\documentclass[10pt,twocolumn,english,prd,superscriptaddress,nofootinbib,preprintnumbers,showpacs,floatfix]{revtex4-2}

\usepackage{graphicx,color}
\usepackage{amsmath}
\usepackage{amssymb}
\usepackage{hyperref}
\usepackage[utf8]{inputenc}
\usepackage[english]{babel}
\usepackage{epsfig}
\usepackage{subfigure}
\usepackage{wasysym}
\usepackage{color,xcolor}
\usepackage{amsmath}
\usepackage{bm}
\usepackage{epsfig}
\usepackage{amsfonts}
\usepackage{dcolumn}
\usepackage{float}

\usepackage{makecell} 
\usepackage{comment}

\usepackage{cases}

\hypersetup{colorlinks=true, linkcolor=blue, citecolor=green}

\usepackage{dcolumn}
\usepackage{bm}
\usepackage{ifpdf}
\usepackage{hyperref}
\usepackage{float}
\usepackage{bm}
\usepackage{xcolor,color,graphicx,graphics}
\usepackage[OT1]{fontenc}
\usepackage{latexsym,amssymb,amsmath,amsfonts}
\usepackage{makeidx}
\usepackage{epsfig}
\usepackage{epstopdf}
\usepackage{mathrsfs}
\hypersetup{colorlinks=true, linkcolor=blue, citecolor=green}
\usepackage{enumerate}
\usepackage{xcolor}
 \usepackage{multirow}

\usepackage{tikz}

\newcommand{\orcidicon}{%
	\begin{tikzpicture}
	\draw[lime, fill=lime] (0,0) 
		circle [radius=0.16] 
		node[white] {{\fontfamily{qag}\selectfont \tiny ID}};
	\draw[white, fill=white] (-0.0625,0.095) 
		circle [radius=0.007];
	\end{tikzpicture}	\hspace{-2mm}
}
\newcommand\orcidLuiz{{\href{https://orcid.org/0009-0004-1652-1393}{\orcidicon}}}
\newcommand\orcidFrancisco{{\href{https://orcid.org/0000-0002-9388-8373}{\orcidicon}}}
\newcommand\orcidManuel{{\href{https://orcid.org/0000-0001-8586-0285}{\orcidicon}}}
\begin{document}

\title{Lorentz-violating signatures in quasi-periodic oscillations from a magnetised Kalb–Ramond black hole}

\author{Luiz F. G. Rodrigues\orcidLuiz\!\!} \email{luizrodrigues.ufpa@gmail.com}
\affiliation{Programa de P\'{o}s-Gradua\c{c}\~{a}o em F\'{i}sica, Universidade Federal do Par\'{a}, 66075-110, Belém, Par\'{a}, Brazil}

\author{Francisco S. N. Lobo\orcidFrancisco\!\!} \email{fslobo@ciencias.ulisboa.pt}
\affiliation{Instituto de Astrof\'{i}sica e Ci\^{e}ncias do Espa\c{c}o, Faculdade de Ci\^{e}ncias da Universidade de Lisboa, Edifício C8, Campo Grande, P-1749-016 Lisbon, Portugal}
\affiliation{Departamento de F\'{i}sica, Faculdade de Ci\^{e}ncias da Universidade de Lisboa, Edif\'{i}cio C8, Campo Grande, P-1749-016 Lisbon, Portugal}

\author{Manuel E. Rodrigues\orcidManuel\!\!}
\email{esialg@gmail.com}
\affiliation{Programa de P\'{o}s-Gradua\c{c}\~{a}o em F\'{i}sica, Universidade Federal do Par\'{a}, 66075-110, Belém, Par\'{a}, Brazil}
\affiliation{Faculdade de Ci\^{e}ncias Exatas e Tecnologia, 
Universidade Federal do Par\'{a}\\
Campus Universit\'{a}rio de Abaetetuba, 68440-000, Abaetetuba, Par\'{a}, 
Brazil}

\date{\today}

\begin{abstract}
We investigate the dynamics of charged particles around a Schwarzschild-like black hole sourced by a Kalb--Ramond field and immersed in a uniform external magnetic field. The Kalb--Ramond field introduces a Lorentz violation parameter $l$ that modifies the spacetime geometry, while the magnetic field profoundly influences the trajectories through the Lorentz force, leading to a rich variety of orbital behaviours including curled (epicyclic) motion and dramatic transitions between distinct energy boundary configurations. We derive the full equations of motion, the effective potential, and the fundamental frequencies of quasi-periodic oscillations, and perform a comprehensive Monte Carlo Markov Chain (MCMC) analysis using observational data from the microquasars GRO 1655-40, XTE 1550-564, and GRS 1915+105. The pure Schwarzschild model is statistically ruled out for all three sources. For GRO 1655-40 and XTE 1550-564, only the combined effect of the magnetic field and Lorentz violation yields statistically robust models, with best-fit values $\mathcal{B}\sim 0.03$--$0.04$ and $l\sim 0.08$--$0.10$. Remarkably, for GRS 1915+105---the most massive object in our sample---the Lorentz violation parameter alone is sufficient to model the QPO frequencies, yielding an optimal fit with $\Delta AIC = \Delta BIC = 0$. Across all three objects, a clear trend emerges: the required value of $l$ decreases as the mass of the astrophysical object increases, suggesting a mass-dependent scaling of the Kalb--Ramond parameter. These findings establish the combined KR and magnetic field framework as a viable and statistically robust scenario for modeling QPOs in microquasars, and indicate that Lorentz-violating modifications to gravity may leave observable imprints in the strong-field regime, offering new avenues for testing quantum-gravity phenomenology with current and next-generation X-ray missions.
\end{abstract}

\maketitle
\def\HMS{{\scriptscriptstyle{\rm HMS}}}

\section{Introduction}\label{sec1}

The possibility that Lorentz symmetry (a cornerstone of both quantum field theory and general relativity) may be violated or deformed at high energies has attracted sustained attention over the past decades. From a theoretical standpoint, several approaches to quantum gravity suggest that Lorentz invariance may not be an exact symmetry, but rather an emergent symmetry that is valid only at low energies. Early motivations arise in string theory, where spontaneous Lorentz symmetry breaking can occur through non-zero vacuum expectation values of tensor fields \cite{KosteleckySamuel1989}. Similarly, loop quantum gravity, non-commutative geometry, and effective field theory frameworks allow for Planck-scale suppressed deviations from Lorentz invariance \cite{AmelinoCamelia1998,Mattingly2005}. The Standard-Model Extension (SME), developed systematically by Kostelecký and collaborators, provides a comprehensive effective field theory framework to parametrise such violations across all sectors of physics \cite{ColladayKostelecky1998,Kostelecky2004}. These developments offer not only conceptual insights into ultraviolet completions of gravity but also a phenomenological arena in which deviations from Lorentz symmetry can be constrained experimentally \cite{Abe2015}.

In gravitational and cosmological contexts, Lorentz violation acquires additional richness. Modified gravity models incorporating preferred vector or tensor fields such as Einstein-Æther theories and Bumblebee models lead to deviations from general relativity that can affect black hole solutions, cosmological expansion, and gravitational wave propagation \cite{Bertolami2005}. In cosmology, Lorentz-violating fields can contribute to inflationary dynamics, dark energy models, or anisotropic stress, potentially leaving imprints in the cosmic microwave background and large-scale structure \cite{Kanno2008}. These scenarios provide a fertile interface between high-energy theory and observational cosmology.

Within this broader landscape, antisymmetric tensor fields play a particularly important role. The Kalb-Ramond (KR) field, originally introduced in the context of string theory \cite{KalbRamond1974}, is a rank-2 antisymmetric tensor field $B_{{\mu}{\nu}}$ whose field strength $H_{{\mu}{\nu}{\rho}}$ generalises the electromagnetic field tensor. Historically, the KR field emerged as a natural excitation mode of closed strings, coupling to string world-sheets analogously to how the electromagnetic potential couples to point particles. Its gauge symmetry and topological features have made it a central object in studies of duality, torsion, and axion-like physics. In gravitational settings, the KR field has been associated with spacetime torsion and has been explored in black hole and cosmological solutions, where it can modify singularity structure, thermodynamics, and propagation of fields \cite{Ednaldo2024,EdnaldoJosé2024,Ednaldo2025,Daniela2025,DanielaEdnaldo2025}. The resulting black hole solution is described by a Schwarzschild-like metric with a Lorentz violation parameter $l$, which controls the deviation from general relativity. Constraints from Solar System tests and black hole shadows restrict $l$ to a narrow range~\cite{Ednaldo2024}, and in this work we adopt the values $l=0$ (Schwarzschild limit), a maximum value of $l=0.189785$, and a minimum value of $l=-0.700225$ as representative references for our trajectory simulations and frequency calculations.

Parallel to the KR field, Bumblebee models describe spontaneous Lorentz symmetry breaking through a vector field $B^{\mu}$ acquiring a non-zero vacuum expectation value~\cite{KosteleckySamuel1989,Samuel1989,Bluhm2005,Capelo2015,Yang2023}. A representative selection of works includes: the foundational SME formulation \cite{ColladayKostelecky1998,Kellie2021,Araújo2026}, which systematises Lorentz violation in effective field theory; studies of Bumblebee gravity showing modified black hole metrics and constraints from astrophysical observations \cite{Casana2018,CarlosMarcos2025,Lessa2025,ZhuXu2025,Lai2026}; and investigations combining antisymmetric tensor fields with Lorentz-violating backgrounds, demonstrating rich phenomenology including modified dispersion relations and birefringence effects~\cite{MalufNeves2011}. Collectively, these works indicate that tensor and vector fields associated with Lorentz violation can produce observable signatures in both strong and weak gravitational regimes \cite{Carroll2004}.

A particularly promising observational window into strong-field gravity is provided by quasi-periodic oscillations (QPOs) observed in X-ray binaries and active galactic nuclei \cite{Psaltis2008,Dasgupta2025,Ahmed2026}. QPOs are quasi-periodic modulations in the X-ray flux emitted by accreting compact objects, believed to originate from oscillatory processes in the inner accretion disc close to the black hole or neutron star \cite{Bambi2012,Falanga2015,Wang2022,Liu2023,Sanjar2023,Boshkayev2023,Jumaniyozov2024,Guo2025,Jumaniyozov2025,Hazarika2025,Hazarika2026}. Historically, their discovery dates back to observations with satellites such as \textit{EXOSAT} and \textit{RXTE}, which revealed both low-frequency (LFQPOs) and high-frequency QPOs (HFQPOs) \cite{StellaVietri1998}, resonance models involving epicyclic frequencies \cite{AbramowiczKluzniak2001,Stuchlik2008,Stuchlik2013,Ditta2024}, and diskoseismology approaches. Observationally, HFQPOs are particularly significant as their frequencies are comparable to orbital frequencies near the innermost stable circular orbit (ISCO), making them sensitive probes of the spacetime geometry. Extensive reviews \cite{Belloni2012} have highlighted the potential of QPOs to test general relativity and alternative theories of gravity.

Another key aspect of realistic astrophysical environments is the presence of external magnetic fields \cite{FrolovShoom2010,Zahrani2013,Kolos2015,Stuchlik2016,Tursunov2016,Gallego2020,Baker2023}. The study of black holes immersed in magnetic fields dates back to the seminal work of Wald \cite{Wald1974}, who demonstrated that a Kerr black hole placed in a uniform magnetic field acquires an induced electric charge. This configuration has become a cornerstone in the study of black hole electrodynamics and magnetohydrodynamics in curved spacetime. In the framework of general relativistic magnetohydrodynamics (GRMHD), magnetic fields play a crucial role in accretion processes, jet formation, and energy extraction mechanisms such as the Blandford-Znajek process \cite{BlandfordZnajek1977}. Observational evidence for strong magnetic fields near black holes is provided by polarimetric measurements and imaging by the Event Horizon Telescope (EHT), which reveal magnetised plasma structures in the vicinity of supermassive black holes \cite{EHT2021}. The interplay between black holes and external fields has led to a variety of important results, including modifications of particle orbits, shifts in ISCO radii, changes in QPO frequencies, and alterations in shadow structure. Studies of magnetised black holes have shown that external fields can significantly influence accretion dynamics and observable signatures \cite{AlievOzdemir2002}. When combined with Lorentz-violating fields or additional tensor degrees of freedom, these effects can become even more pronounced, offering potential avenues for observational constraints.

In this work, we investigate a Schwarzschild-like solution sourced by a Kalb-Ramond field and consider its embedding in an external magnetic field aligned along the $z$-axis. Our aim is to explore how the combined effects of antisymmetric tensor fields and external electromagnetic environments modify the spacetime geometry and its astrophysical signatures. In particular, we analyse the implications for particle dynamics and QPO observables, providing a bridge between fundamental field-theoretic modifications and measurable astrophysical phenomena. We perform a Monte Carlo Markov Chain (MCMC) statistical analysis to constrain the model parameters using observed QPO frequencies from the microquasars GRO 1655-40, XTE 1550-564, and GRS 1915+105, demonstrating that the combined KR and magnetic field framework provides statistically robust fits to the data.

This paper is organized as follows: in Section \ref{sec2}, we present the metric and the equations of motion, along with the effective potential and its properties. In Section \ref{sec3}, we discuss the trajectories of charged particles, presenting several examples, with an emphasis on curled trajectories. In Section \ref{sec4}, we present the frequencies of QPOs and statistical analyses of astrophysical objects associated with Gaussian distributions in order to determine the best models for the quasi-periodic oscillations. We conclude in Section \ref{sec5} with a summary of our findings and directions for future research. In this paper, we will use geometrized units $G=c=1$. When studying the frequencies of QPOs, we will revert to SI units.

\section{Kalb-Ramond black hole metric and electromagnetic field}\label{sec2}

\subsection{Metric and four-vector potential}

The metric of a Kalb-Ramond black hole, obtained from the modified Einstein equations sourced by the KR field~\cite{Ednaldo2024}, is given by
\begin{equation}
	ds^{2}=-f(r)dt^{2}+f^{-1}(r)dr^{2}+r^{2}d{\theta}^{2}+r^{2}\sin^{2}{\theta}d{\phi}^{2},
	\label{ds2}
\end{equation}
where the function $f(r)$ is
\begin{equation}
	f(r)=\frac{1}{1-l}-\frac{2M}{r}.
	\label{fr}
\end{equation}
The term $l$ represents the Lorentz violation parameter, which corresponds to the Kalb-Ramond field.  For small $l$, $f(r)\approx 1+l-2M/r$, and the Schwarzschild solution is recovered in the limit $l\to 0$.

Since the Kalb-Ramond spacetime preserves stationary and axial symmetry, we have the existence of space-type and time-type Killing vectors. These vectors satisfy the Killing equation
\begin{equation}
	{\xi}_{{\alpha},{\beta}}+{\xi}_{{\beta},{\alpha}}=0.
	\label{xi}
\end{equation}
Thus, the four-vector potential $A^{\mu}$ for the electromagnetic field will have a solution of the form~\cite{Wald1984book}
\begin{equation}
	A^{\mu}=C_{1}{\xi}^{\mu}_{(t)}+C_{2}{\xi}^{\mu}_{({\phi})}.
	\label{Amu}
\end{equation}

From here on, we will consider the existence of a uniform magnetic field whose source is located at infinite distances in space. This field has intensity $B$ and is oriented perpendicularly to the equatorial plane of the black hole's event horizon. Therefore the potential four-vector becomes
\begin{equation}
	A^{\alpha}=\frac{B}{2}{\xi}^{\alpha}_{({\phi})}.
	\label{Aalpha}
\end{equation}

The Killing vector associated with azimuthal symmetry can be taken as the basis vector for the $\phi$-coordinate, and thus ${\xi}_{({\phi})}={\partial}/{\partial}{\phi}$, which generates rotations about the axis of symmetry. This implies that only one covariant component of the four-vector potential is non-zero, which means that the electromagnetic field potential is
\begin{equation}
	A_{\phi}=\frac{B}{2}g_{{\phi}{\phi}}=\frac{B}{2}r^{2}\sin^{2}{\theta}.
	\label{Aphi}
\end{equation}

\subsection{Lagrangian, equations of motion and conserved quantities}

The Lagrangian of a charged particle in Kalb-Ramond spacetime, in the presence of an external magnetic field, is given by
\begin{equation}
	\begin{split}
		\mathcal{L}=\frac{1}{2}m\left[-f(r)\dot{t}^{2}+f^{-1}(r)\dot{r}^{2}+r^{2}\dot{\theta}^{2}+r^{2}\sin^{2}{\theta}\dot{\phi}^{2}\right] \\
		+\frac{qB}{2}r^{2}\sin^{2}{\theta}\dot{\phi}.
	\end{split}
	\label{lagrangeana}
\end{equation}
We can determine the equations of motion from the Euler-Lagrange equations
\begin{equation}
	\frac{d}{d{\tau}}\left(\frac{{\partial}\mathcal{L}}{{\partial}\dot{x}^{\mu}}\right)-\frac{{\partial}\mathcal{L}}{{\partial}x^{\mu}}=0.
	\label{eulerlag}
\end{equation}
Since the Lagrangian does not explicitly depend on the $t$-coordinate, for the first case, we have
\begin{equation}
	\frac{d}{d{\tau}}\left(-mf(r)\dot{t}\right)=0.
	\label{dt}
\end{equation}
Thus, we have the first constant of motion, which gives us the specific energy of the particle
\begin{equation}
	\mathcal{E}=f(r)\dot{t}.
	\label{E}
\end{equation}
On the other hand, for the azimuthal coordinate, since the Lagrangian does not explicitly depend on the $\phi$-coordinate, then
\begin{equation}
	\frac{d}{d{\tau}}\left(mr^{2}\sin^{2}{\theta}\dot{\phi}+\frac{qB}{2}r^{2}\sin^{2}{\theta}\right)=0,
	\label{dphi}
\end{equation}
so we obtain the other constant of motion, associated with the axial angular momentum of the particle
\begin{equation}
	L=mr^2\sin^{2}{\theta}\left(\dot{\phi}+\frac{qB}{2m}\right).
	\label{L}
\end{equation}
For the equation of motion in the $\theta$-coordinate, the Euler–Lagrange equations lead to
\begin{equation}
	\frac{d}{d{\tau}}\left(\frac{{\partial}\mathcal{L}}{{\partial}\dot{\theta}}\right)=2mr\dot{r}\dot{\theta}+mr^{2}\ddot{\theta},
	\label{dtheta1}
\end{equation}
\begin{equation}
	\frac{{\partial}\mathcal{L}}{{\partial}{\theta}}=\frac{1}{2}mr^{2}\sin{2{\theta}}\dot{\phi}^{2}+\frac{qB}{2}r^{2}\sin{2{\theta}}\dot{\phi}.
	\label{dtheta2}
\end{equation}
Combining equations \eqref{dtheta1} and \eqref{dtheta2}, we then obtain
\begin{equation}
	\ddot{\theta}=-\frac{2}{r}\dot{r}\dot{\theta}+\frac{1}{2}\sin{2{\theta}}\dot{\phi}^{2}+\mathcal{B}\sin{2{\theta}}\dot{\phi},
	\label{thetaddot}
\end{equation}
where we define the magnetic field parameter as~\cite{FrolovShoom2010}
\begin{equation}
	\mathcal{B}=\frac{qB}{2m}.
	\label{Bcalig}
\end{equation}
Finally, let us determine the equation of motion for the $r$-coordinate. From the Euler-Lagrange equations
\begin{equation}
	\frac{d}{d{\tau}}\left(\frac{{\partial}\mathcal{L}}{{\partial}\dot{r}}\right)=mf^{-1}(r)\ddot{r}-m\frac{f^{\prime}(r)}{f^{2}(r)}\dot{r}^{2},
	\label{dr1}
\end{equation}
	\begin{eqnarray}
		\frac{{\partial}\mathcal{L}}{{\partial}r}=-\frac{m}{2}f^{\prime}(r)\dot{t}^{2}-\frac{m}{2}\frac{f^{\prime}(r)}{f^{2}(r)}\dot{r}^{2}+mr\dot{\theta}^{2}
			\nonumber \\
		+mr\sin^{2}{\theta}\dot{\phi}^{2}+qBr\sin^{2}{\theta}\dot{\phi}.
		\label{dr2}
	\end{eqnarray}

Thus, combining equations \eqref{dr1} and \eqref{dr2}, we have that the equation of motion for the radial coordinate is expressed by
	\begin{eqnarray}
		\ddot{r}=\frac{1}{2}\frac{f^{\prime}(r)}{f(r)}\dot{r}^{2}-\frac{1}{2}f(r)f^{\prime}(r)\dot{t}^{2}+rf(r)\dot{\theta}^{2}
			\nonumber \\
		+rf(r)\sin^{2}{\theta}\left(\dot{\phi}^{2}+2\mathcal{B}\dot{\phi}\right).
		\label{rddot}
	\end{eqnarray}

We now introduce the specific angular momentum
\begin{equation}
	\mathcal{L}=\frac{L}{m},
	\label{Lcaig}
\end{equation}
together with the specific energy already defined in Eq.~\eqref{E}.
With these definitions and that of equation \eqref{Bcalig}, the equations of motion can be rewritten as
\begin{equation}
	\dot{t}=\frac{\mathcal{E}}{f(r)},
	\label{tdot}
\end{equation}
\begin{equation}
	\dot{\phi}=\frac{\mathcal{L}}{r^{2}\sin^{2}{\theta}}-\mathcal{B},
	\label{phidot}
\end{equation}
\begin{equation}
	\ddot{\theta}=-\frac{2}{r}\dot{r}\dot{\theta}+\frac{1}{2}\sin{2{\theta}}\left[\left(\frac{\mathcal{L}}{r^{2}\sin^{2}{\theta}}\right)^{2}-\mathcal{B}^{2}\right],
	\label{thetaddot2}
\end{equation}
	\begin{eqnarray}
		\ddot{r}&=&\frac{1}{2}\frac{f^{\prime}(r)}{f(r)}\dot{r}^{2}-\frac{1}{2}\frac{f^{\prime}(r)}{f(r)}\mathcal{E}^{2}+rf(r)\dot{\theta}^{2}
			\nonumber \\
		&&+rf(r)\sin^{2}{\theta}\left[\left(\frac{\mathcal{L}}{r^{2}\sin^{2}{\theta}}\right)^{2}-\mathcal{B}^{2}\right].
		\label{rddot2}
	\end{eqnarray}

\subsection{Effective potential and stable orbits}

For the signature we are using, the normalization of the quad-speed is, in natural units
\begin{equation}
g_{{\mu}{\nu}}u^{\mu}u^{\nu}=-1.
\label{quadveloc}
\end{equation}

Taking the Lagrangian of equation \eqref{lagrangeana} and considering the fact that the motion of the particles occurs in a fixed plane, we have that $\dot{\theta}=0$, and thus we obtain
\begin{equation}
\dot{r}^{2}=\mathcal{E}^{2}-f(r)\left[1+\left(\frac{\mathcal{L}}{r\sin{\theta}}-\mathcal{B}r\sin{\theta}\right)^{2}\right].
\label{rdot2}
\end{equation}

Thus, we can identify the effective potential, given by
\begin{equation}
V_{\mathrm{eff}}(r,{\theta})=f(r)\left[1+\left(\frac{\mathcal{L}}{r\sin{\theta}}-\mathcal{B}r\sin{\theta}\right)^{2}\right].
\label{Veff}
\end{equation}

From here, we can perform a signal analysis on the specific angular momentum $\mathcal{L}$ and the magnetic field parameter $\mathcal{B}$, based on the term in parentheses, which represents the potential of the central force. This analysis is important for predicting the behavior of charged particles around the black hole, which could allow them to escape to infinity, be absorbed, or even remain confined to a stable orbital region. The equation for the effective potential \eqref{Veff} reveals a symmetry between the specific angular momentum and the magnetic parameter. This symmetry is expressed by $(\mathcal{L},\mathcal{B})\longleftrightarrow(-\mathcal{L},-\mathcal{B})$. We can divide this symmetry into two cases: the \textit{minus configuration} and the \textit{plus configuration}.

\begin{itemize}
\item Minus configuration: in this case, the magnetic parameter and the angular momentum have opposite signs, so that the Lorentz force is attractive and draws the charged particle along the $z$-axis towards the black hole.

\item Plus configuration: here, the magnetic field and the angular momentum have the same signs; therefore, the Lorentz force will be repulsive, pushing the charged particle away from the black hole.
\end{itemize}

The motion of the charged particle will be constrained by energy boundaries, given by the expression
\begin{equation}
\mathcal{E}^{2}=V_{\mathrm{eff}}(r,{\theta}).
\label{E2}
\end{equation}

\begin{figure*}[t]
\centering
\subfigure[\label{VxB0}]{\includegraphics[width=0.32\textwidth]{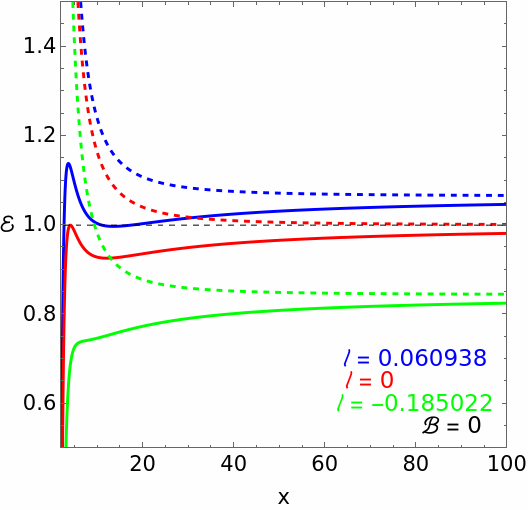}}
\hfill
\subfigure[\label{VxB-}]{\includegraphics[width=0.32\textwidth]{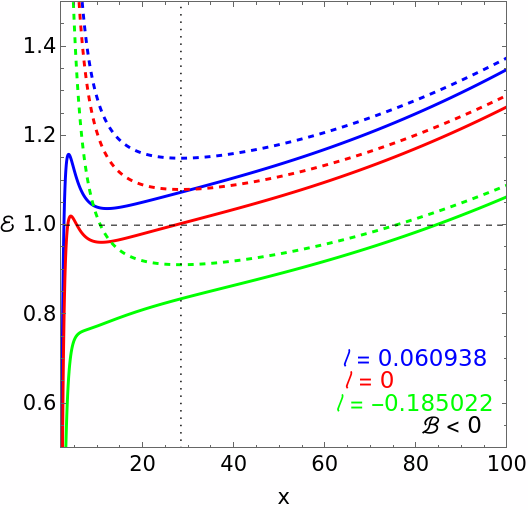}}
\hfill
\subfigure[\label{VxB+}]{\includegraphics[width=0.32\textwidth]{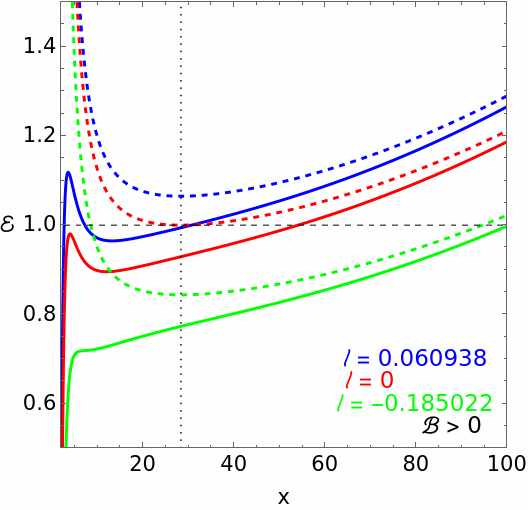}}
\caption{Sections ($y=0,z=\mathrm{constant}$) of the effective potential $V_{\mathrm{eff}}$ taken in the equatorial plane ${\theta}={\pi}/2$ when $z=0$ (solid curves) and at spatial infinity $z\rightarrow\infty$ (dashed curves) for a charged particle moving around a Kalb-Ramond black hole immersed in a uniform external magnetic field. We compare the cases with and without the magnetic field: $\mathcal{B}=0$ (a), $\mathcal{B}=-0.005$ (b), and $\mathcal{B}=0.005$ (c). For these three cases, the angular momentum $\mathcal{L}=4$ was used. We also compare the presence of the Lorentz violation parameter with the red curves ($l=0$), blue curves ($l>0$), and green curves ($l<0$). The minimum of the effective potential is located at $r=2\sqrt{20}$ (dotted vertical line). As we can see, the magnetic field is a decisive factor for both small and large values of the $x$-coordinate. On the other hand, the presence of the Lorentz violation parameter has a certain influence on the strength of the effective potential.}
\label{Vx}
\end{figure*}

From this, we can examine certain characteristics of the effective potential, which enable us to determine the general properties of the charged particle’s motion. In Fig. \ref{Vx}, we can see the behavior of the effective potential along the $x$-axis in the absence and presence of a magnetic field, as well as with the Lorentz violation parameter.

The critical points of the effective potential, which are candidates for local maxima or minima, can be determined using the following expressions
\begin{equation}
\begin{split}
{\partial}_{r}V_{\mathrm{eff}}(r,{\theta})=0, \\
{\partial}_{\theta}V_{\mathrm{eff}}(r,{\theta})=0.
\end{split}
\label{drdtheta}
\end{equation}

The first derivative of the effective potential with respect to $\theta$ gives us the value ${\theta}={\pi}/2$. This means that the critical points of the effective potential lie in the equatorial plane. Therefore, the trajectories of the charged particles must occur primarily in this region.

On the other hand, the first derivative with respect to the $r$-coordinate yields the following fifth-order polynomial
\begin{equation}
\mathcal{L}^{2}\left(\frac{r}{1-l}-3\right)+2\mathcal{L}\mathcal{B}r^{2}-\mathcal{B}^{2}r^{4}\left(\frac{r}{1-l}-1\right)-r^{2}=0.
\label{dVeffdr}
\end{equation}

The real roots of equation \eqref{dVeffdr} above the black hole’s event horizon give us the maximum, minimum, and inflection points of the effective potential. From these extreme points, we can determine the positions of stable equilibrium, indicated by the minima, and of unstable equilibrium, indicated by the maxima. On the other hand, the inflection points give us the circular orbits that are marginally stable.

We can see that equation \eqref{dVeffdr} is quadratic with respect to the specific angular momentum, so we can determine the circular orbits from the following expression
\begin{equation}
\mathcal{L}_{E{\pm}}(r)=\frac{-(1-l)\mathcal{B}r^{2}{\pm}(1-l)rA}{r-3(1-l)},
\label{LE+-}
\end{equation}
where
\begin{equation}
A(r)=\sqrt{\frac{\left[\left(1-l\right)\left(r-3\left(1-l\right)\right)\right]+\mathcal{B}^{2}r^{2}\left[r-2\left(1-l\right)\right]^{2}}{\left(1-l\right)^{2}}}.
\label{Ar}
\end{equation}

The positive branch of the equation \eqref{LE+-}, for which the solution is $\mathcal{L}_{E+}>0$, can occur for the entire region $r>2$ and thus determines both the stable and unstable circular orbits. An interesting fact is that, in the case of the plus configuration, where $\mathcal{B}>0$, both solutions $\mathcal{L}_{E{\pm}}>0$ may exist in the region $2<r<3$. On the other hand, the negative branch of the equation \eqref{LE+-}, whose solution is $\mathcal{L}_{E-}$, gives us only the maximum of the effective potential. By taking the effective potential and setting ${\partial}_{r}^{2}V_{\mathrm{eff}}=0$, we obtain an expression that yields the innermost stable circular orbit (ISCO).

We know that the motion of an uncharged test particle is confined to the equatorial plane in Schwarzschild spacetime. However, for the motion of a charged particle in the presence of a uniform magnetic field and Lorentz violation, the motion will occur in the $r$ direction close to the equatorial plane. This is because the term $\mathcal{B}^{2}r^{2}$ present in the effective potential grows indefinitely as $r\rightarrow\infty$. Furthermore, the energy boundary of equation \eqref{E2} for the case of a charged particle allows for motion in the polar direction of the black hole’s spacetime. This opens up the possibility of particles escaping to infinity along the $z$-direction. The minimum energy required for a charged particle to escape to infinity is
\begin{equation}
\mathcal{E}\geq\mathcal{E}_{\mathrm{flat(min)}}=\left\lbrace\begin{matrix}
\sqrt{\frac{1}{1-l}} & \mathrm{for} & \mathcal{B}\geq0, \\
\sqrt{\frac{1-4\mathcal{B}\mathcal{L}}{1-l}} & \mathrm{for} & \mathcal{B}<0,
\end{matrix}\right.
\label{Eflat}
\end{equation}

If the energy boundary of equation \eqref{E2} forms a closed curve, then the motion of the charged particle becomes confined, such that its trajectory is restricted to a toroidal region governed by the effective potential. The condition for obtaining this type of region, confined by the energy boundary, is related to the minimum of the effective potential, and thus, we have that
\begin{equation}
\mathcal{E}>\mathcal{E}_{\mathrm{flat(min)}}.
\label{Eflat2}
\end{equation}

From this, we can determine the confinement region in which the particle can exist, and thus
\begin{equation}
\mathcal{L}_{\mathrm{L1}}<\mathcal{L}<\mathcal{L}_{\mathrm{L2}}.
\label{L1LL2}
\end{equation}

So, the functions $\mathcal{L}_{\mathrm{L1}}(r)$ and $\mathcal{L}_{\mathrm{L2}}(r)$ define the confinement region, which we can call the `lake'´. These functions are solutions to the following condition
\begin{equation}
\mathcal{E}=\mathcal{E}_{\mathrm{flat(min)}}.
\label{Eflat3}
\end{equation}

Therefore, from the condition of equation \eqref{Eflat3}, we can determine the expressions for the functions $\mathcal{L}_{\mathrm{L1}}(r)$ and $\mathcal{L}_{\mathrm{L2}}(r)$. For the first case, where $\mathcal{B}\geq0$, they are given by
\begin{equation}
\mathcal{L}_{\mathrm{L1,L2}}=\mathcal{B}r^{2}{\pm}\frac{r\sqrt{\left[r-2\left(1-l\right)\right]\left[2-l\left(r+2\right)\right]}}{r-2\left(1-l\right)}.
\label{LL1L2B+}
\end{equation}

If $\mathcal{B}<0$, they become
\begin{equation}
\begin{split}
\mathcal{L}_{\mathrm{L1,L2}}=\frac{-\mathcal{B}r^{2}\left[r+2+2l\left(r+1\right)\right]}{r-2\left(1-l\right)} \\
{\pm}\frac{r\sqrt{\left[2-l\left(r+2\right)\right]\left[r-2\left(1-l\right)+4\mathcal{B}^{2}r^{3}\left(1-l\right)\right]}}{r-2\left(1-l\right)}.
\end{split}
\label{LL1L2B-}
\end{equation}

\begin{figure*}[t]
\centering
\subfigure[\label{LrB0}]{\includegraphics[width=0.32\textwidth]{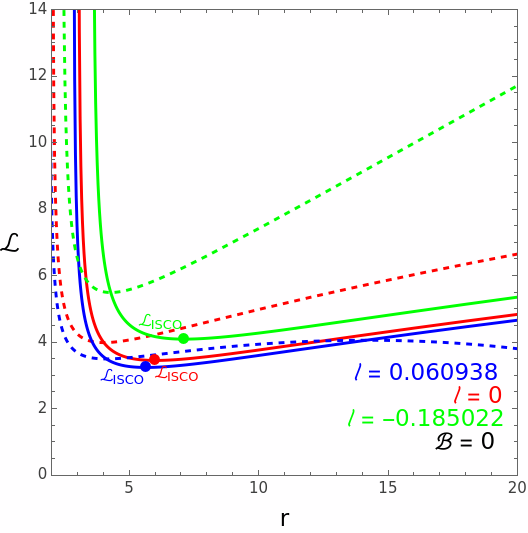}}
\hfill
\subfigure[\label{LrB-}]{\includegraphics[width=0.32\textwidth]{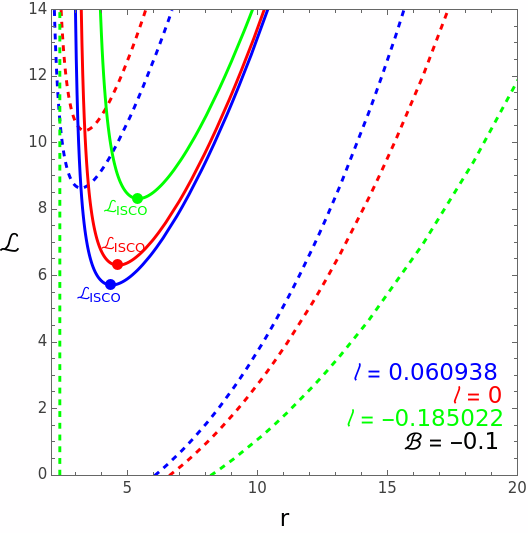}}
\hfill
\subfigure[\label{LrB+}]{\includegraphics[width=0.32\textwidth]{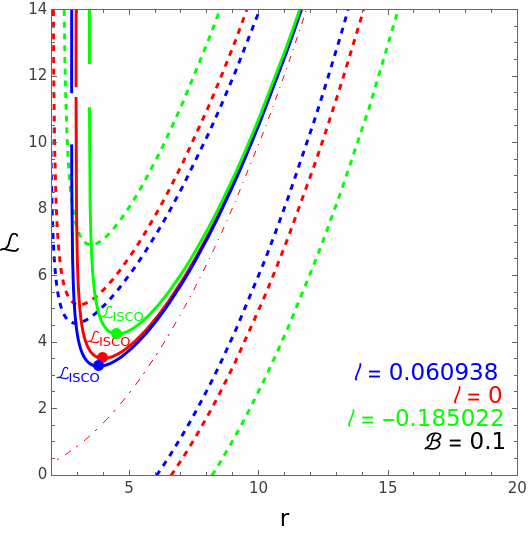}}
\caption{Circular orbits of charged particles in the Kalb–Ramond spacetime. The $\mathcal{L}_{\mathrm{E{\pm}}}(r)$ function (solid curve) determines the local maximum of the effective potential and, consequently, the circular orbits in the equatorial plane. The $\mathcal{L}_{\mathrm{ISCO}}$ point governs the innermost stable circular orbit, such that for $r>r_{\mathrm{ISCO}}$ we have stable orbits, and for $r<r_{\mathrm{ISCO}}$ unstable orbits. The $\mathcal{L}_{\mathrm{L1,L2}}(r)$ functions (dashed curves) determine the region of trapped states. The negative branch of the extreme function $\mathcal{L}_{\mathrm{E-}}(r)$ and the function $\mathcal{L}_{*}(r)$ (dotted-dashed curve) determine the existence of trajectories with rotations, which exist only for $\mathcal{B}>0$. All curves have been plotted for various values of $\mathcal{B}$ and $l$.}
\label{Lr}
\end{figure*}

We can see the behaviour of these $\mathcal{L}_{\mathrm{L1,L2}}$ functions and the $\mathcal{L}_{\mathrm{E}{\pm}}$ solutions for various values of $\mathcal{B}$ and $l$ in Fig. \ref{Lr}.

\begin{figure*}[t]
\centering
\subfigure[\label{LrB0l0}]{\includegraphics[width=0.32\textwidth]{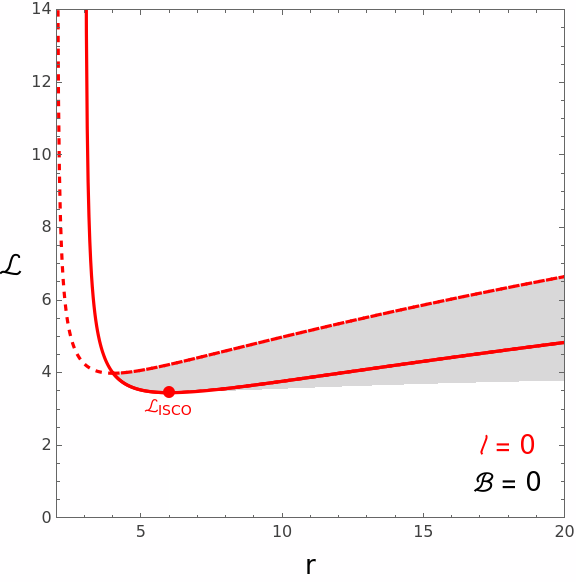}}
\hfill
\subfigure[\label{LrB0l+}]{\includegraphics[width=0.32\textwidth]{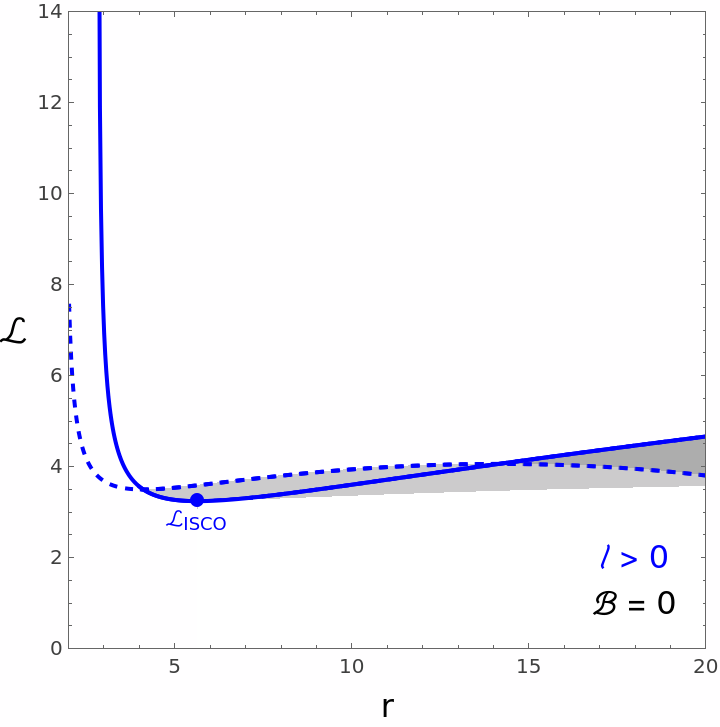}}
\hfill
\subfigure[\label{LrB0l-}]{\includegraphics[width=0.32\textwidth]{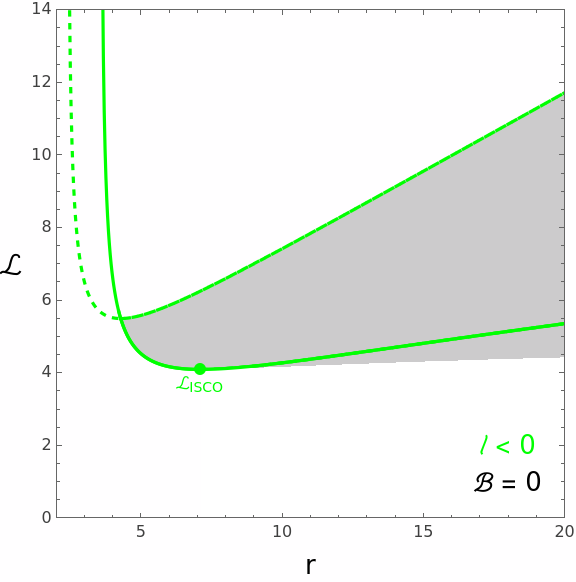}}
\hfill
\subfigure[\label{LrB-l0}]{\includegraphics[width=0.32\textwidth]{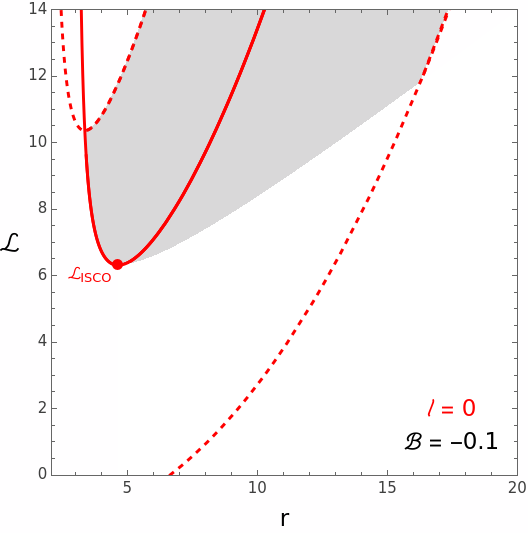}}
\hfill
\subfigure[\label{LrB-l+}]{\includegraphics[width=0.32\textwidth]{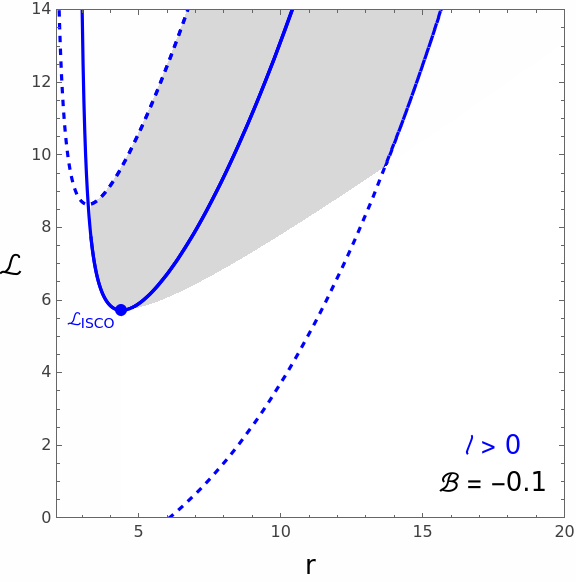}}
\hfill
\subfigure[\label{LrB-l-}]{\includegraphics[width=0.32\textwidth]{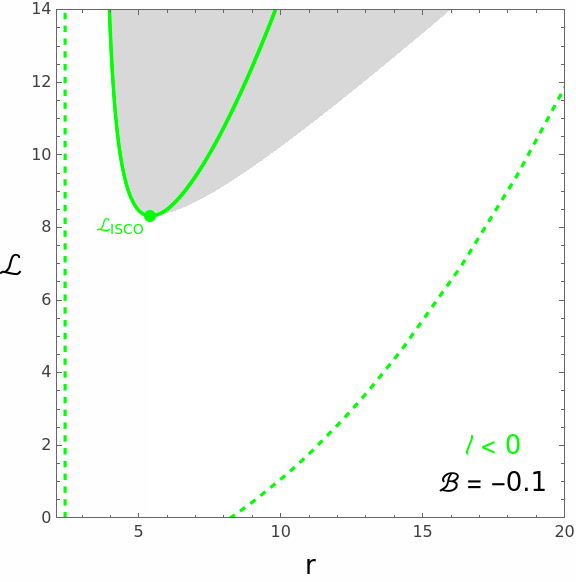}}
\hfill
\subfigure[\label{LrB+l0}]{\includegraphics[width=0.32\textwidth]{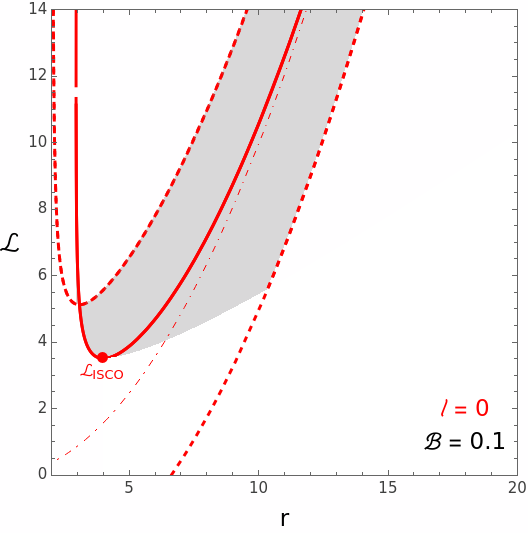}}
\hfill
\subfigure[\label{LrB+l+}]{\includegraphics[width=0.32\textwidth]{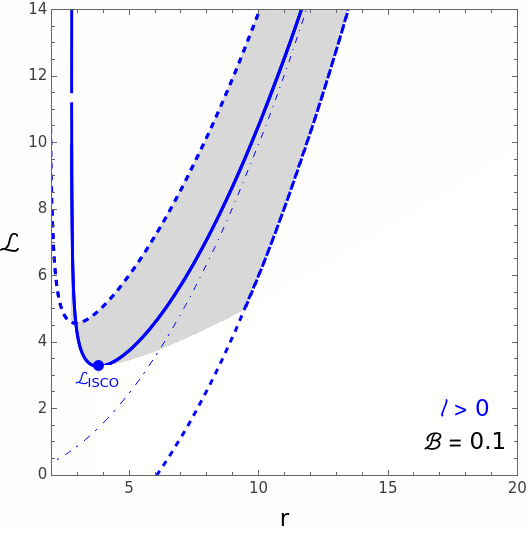}}
\hfill
\subfigure[\label{LrB+l-}]{\includegraphics[width=0.32\textwidth]{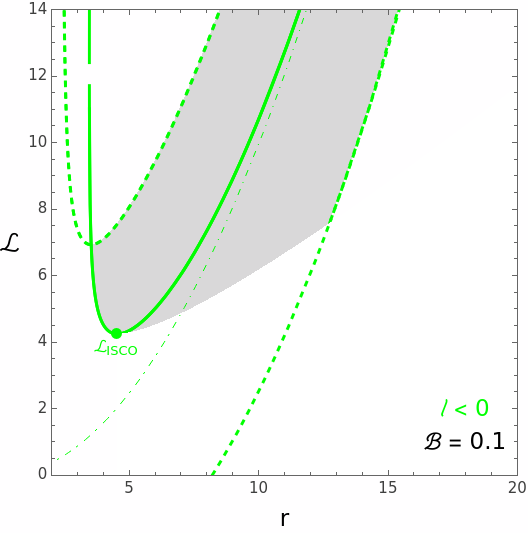}}
\caption{Specific angular momentum as a function of the $r$-coordinate. We have the absence of a magnetic field and Lorentz violation ($\mathcal{B}=0,l=0$), as well as combinations of positive and negative values for these parameters. The shaded region indicates the region of trapped states. For the plots where $\mathcal{B}>0$, there is the possibility of curled trajectories within these regions.}
\label{LrBl}
\end{figure*}

Figure \ref{Lr} illustrates the trapped states of the charged particle trajectories, as well as the conditions under which curled trajectories exist. These regions of trapped states are identified in Figure \ref{LrBl} for each set of values of $\mathcal{B}$ and $l$.

\begin{figure*}[t]
\centering
\subfigure[\label{ELB0}]{\includegraphics[width=0.32\textwidth]{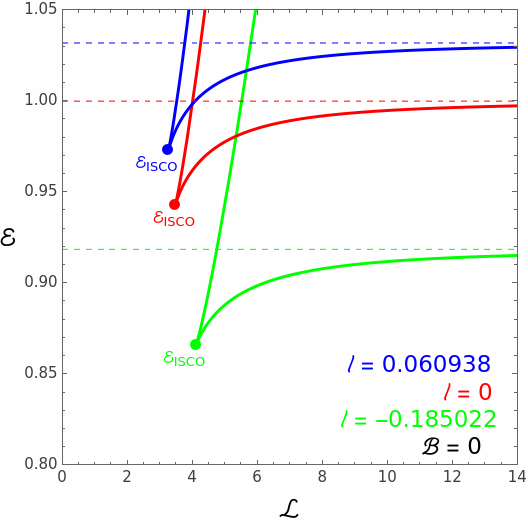}}
\hfill
\subfigure[\label{ELB-}]{\includegraphics[width=0.32\textwidth]{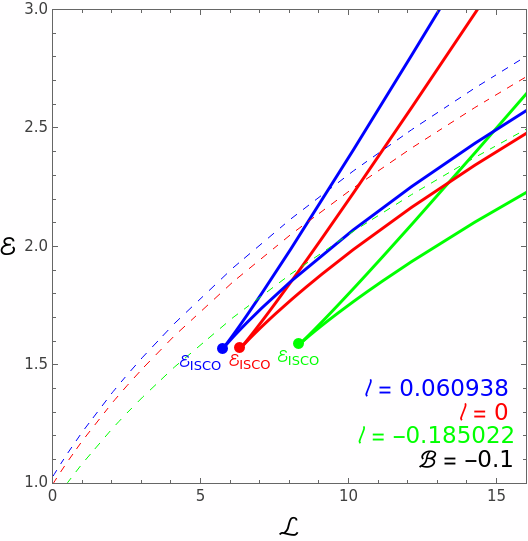}}
\hfill
\subfigure[\label{ELB+}]{\includegraphics[width=0.32\textwidth]{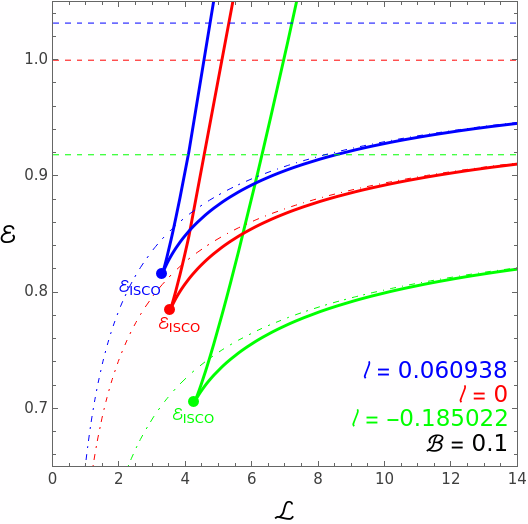}}
\caption{Energies of circular orbits. The local maximum of the effective potential determines the circular orbits in the equatorial plane, which have been expressed as functions of the specific angular momentum. The solid curves correspond to the maximum (curves on the left) and minimum (curves on the right) of the effective potential and to unstable and stable orbits, respectively. The minimum of the effective potential for the flat spacetime $\mathcal{E}_{\mathrm{flat(min)}}$ is represented by the dashed curves, which identify the escape region to infinity. In the case $\mathcal{B}>0$, curly motion is possible for particles with parameters above the $\mathcal{E}_{*}(\mathcal{L})$ function (dotted-dashed curves). The graphs were generated from various combinations of $\mathcal{B}$ and $l$.}
\label{EL}
\end{figure*}

Figure \ref{EL} shows the energy of the circular orbits of charged particles, as determined by the maximum and minimum values of the effective potential, for various values of the magnetic field and the Lorentz violation parameter.

\section{Trajectories of charged particles}\label{sec3}

Now, let us examine the trajectories of charged particles around a Kalb-Ramond black hole and understand how the Lorentz violation parameter and the presence of a magnetic field significantly alter the particles’ paths. For the purposes of analysis, we will compare the trajectories in the Schwarzschild and Kalb-Ramond spacetimes, both of which are immersed in a uniform external magnetic field.

\subsection{General trajectories}

From the conservation of angular momentum, as given by equation \eqref{L}, we find that the equation of motion in the axial coordinate for a charged particle orbiting a Kalb-Ramond black hole immersed in a uniform magnetic field, with respect to the equatorial plane, is given by
\begin{equation}
\dot{\phi}=\frac{\mathcal{L}}{r^{2}}-\mathcal{B}.
\label{phidot2.0}
\end{equation}

In the case of no magnetic field, the right-hand side of the equation \eqref{phidot2.0} is always positive and increases continuously. However, for positive values of the magnetic field parameter, the $\phi$-coordinate may decrease. Thus
\begin{equation}
\mathcal{L}>\mathcal{L}_{*}(r)=\mathcal{B}r^{2}.
\label{L*}
\end{equation}

In the case of energy, we can express this condition as
\begin{equation}
\mathcal{E}>\mathcal{E}_{*}(\mathcal{L})=\sqrt{\frac{1}{1-l}-2\sqrt{\frac{\mathcal{B}}{\mathcal{L}}}}.
\label{E*}
\end{equation}

As the $\phi$-coordinate decreases, the motion of the charged particle becomes epicyclic, with the curled trajectories. The $\mathcal{L}_{*}(r)$ and $\mathcal{E}_{*}(\mathcal{L})$ functions, which are responsible for the curling of the trajectories, are illustrated in Figs. \ref{Lr} and \ref{EL}, respectively.

The motion of charged particles around a black hole embedded in a magnetic field is generally chaotic. In our case, for a Kalb-Ramond black hole, we will observe similar behaviour. However, for the motion of charged particles near the minimum of the effective potential - and consequently near circular orbits - we can observe harmonic and regular motion. Furthermore, trajectories that lie entirely within the equatorial plane will also be regular; chaotic motion arises due to changes in the inclination angle ${\theta}_{0}$ relative to the equatorial plane. However, our focus will be on harmonic or quasi-harmonic trajectories close to the equatorial plane.

\begin{figure}[htpb]
\centering
\includegraphics[width=0.4\textwidth]{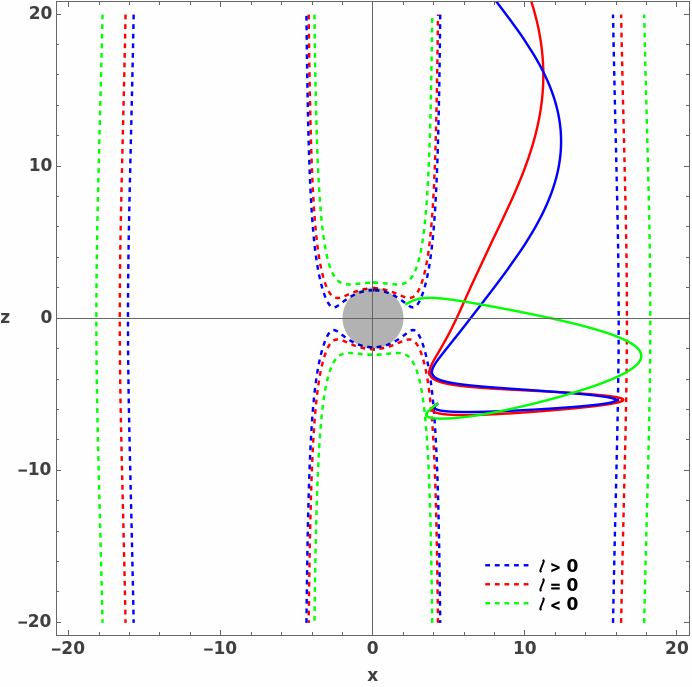}
\caption{Case where there are no inner or outer boundaries. The dashed lines represent $\mathcal{E}^{2}=V_{\mathrm{eff}}$, which delimit the regions of valid motion for the charged particles. The solid curves represent the trajectories of some particles with the parameters $\mathcal{L}=7,\mathcal{E}=1.5,\mathcal{B}=0.1$, and $l=(-0.185022,0.060938)$.}
\label{1caso}
\end{figure}

\begin{figure}[htpb]
\centering
\includegraphics[width=0.4\textwidth]{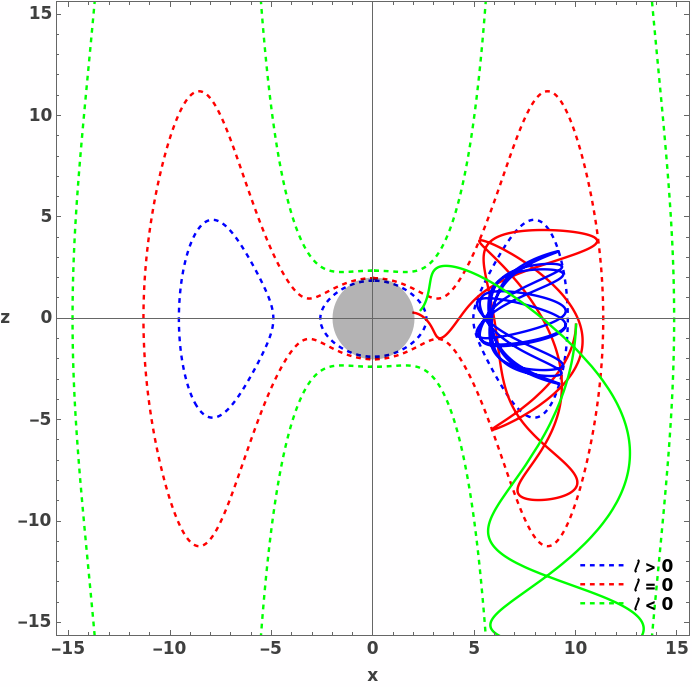}
\caption{In this situation of an outer boundary, with the parameters ($\mathcal{L}=8,\mathcal{E}=1.9,\mathcal{B}=-0.1$, and $l=(-0.185022,0.060938)$), the outer boundary exists only in the Schwarzschild case ($l=0$). For Kalb-Ramond situations, we observe inner and outer boundaries for positive $l$, and no boundaries for negative $l$.}
\label{2caso}
\end{figure}

\begin{figure}[htpb]
\centering
\includegraphics[width=0.4\textwidth]{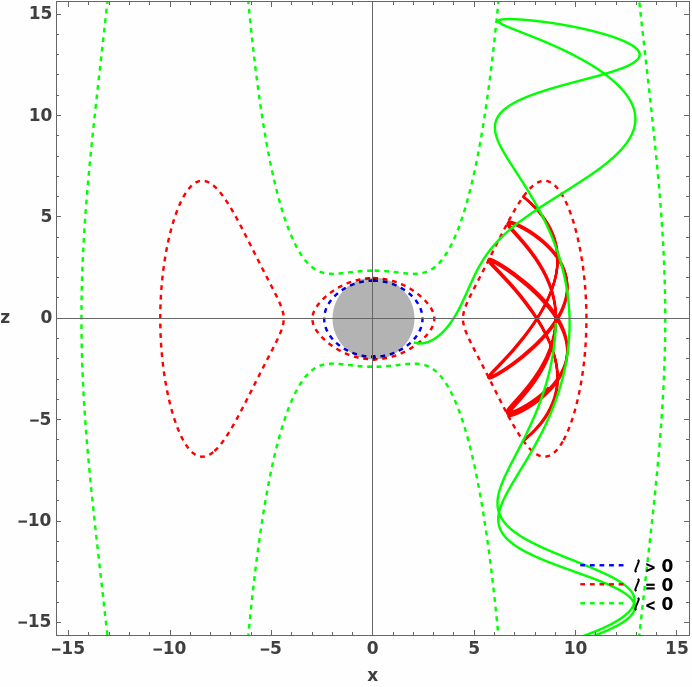}
\caption{Here we have the case of inner and outer borders, with the parameters ($\mathcal{L}=8.5,\mathcal{E}=1.9,\mathcal{B}=-0.1$, and $l=(-0.185022,0.060938)$). Both boundaries only exist in the case of Schwarzschild, where $l=0$. When $l$ is positive, we have only the inner boundary, and when $l$ is negative, there are no boundaries. It is worth noting that, with these parameters, it was not possible to find a physically consistent trajectory for $l>0$.}
\label{3caso}
\end{figure}

\begin{figure}[htpb]
\centering
\includegraphics[width=0.4\textwidth]{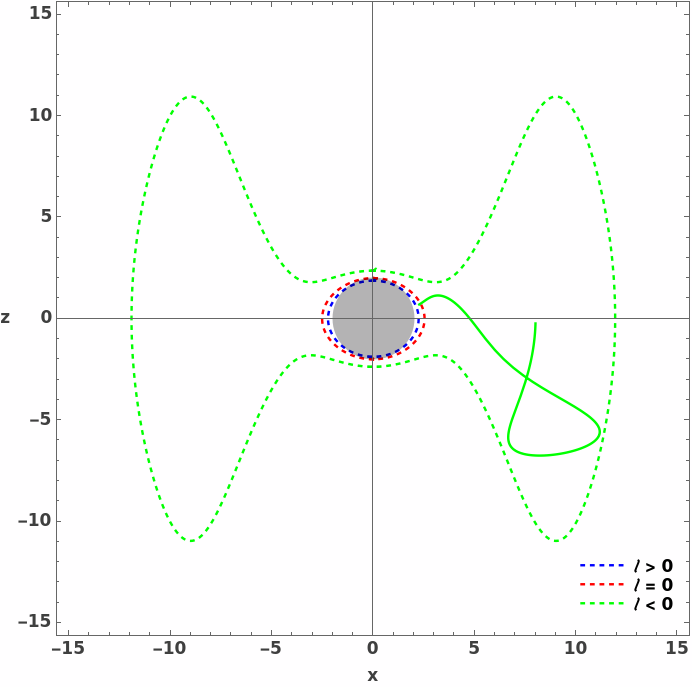}
\caption{Now we have the case where only the inner boundary exists. Given the parameters ($\mathcal{L}=8.5,\mathcal{E}=1.9,\mathcal{B}=-0.1$, and $l=(-0.185022,0.060938)$). We can see that the inner boundary occurs when $l=0$ and $l>0$. When $l$ is negative, an outer boundary appears.}
\label{4caso}
\end{figure}

Based on the behaviour of the effective potential, we can distinguish four types of energy boundaries from equation \eqref{E2}.
\begin{itemize}
\item First case: This corresponds to the absence of any inner or outer boundary. In this situation, the charged particle is liable to be captured by the black hole or to escape into infinity. Figure \ref{1caso}.

\item Second case: in this situation, there is an outer boundary. As a result, the charged particle can be captured by the black hole. Figure \ref{2caso}.

\item Third case: indicates the existence of both inner and outer energy boundaries. In this case, the charged particle becomes trapped in a toroidal region around the black hole. This condition gives rise to the ``lake" potential around the black hole’s event horizon. Figure \ref{3caso}.

\item Fourth case: in this situation, there is only the inner boundary, so the particle cannot be absorbed by the black hole, but it can escape into infinity. Figure \ref{4caso}.
\end{itemize}

We therefore present the four cases of energy boundaries for the motion of charged particles. It should be noted that we have taken the Schwarzschild spacetime as a reference for these cases, without the Lorentz violation parameter. In general, when the Lorentz violation (Kalb-Ramond) is included, the boundaries remain in the same configuration. However, in some cases, we can observe a drastic change in the type of boundary, as illustrated in Figs. \ref{2caso} and \ref{3caso}. Furthermore, as already mentioned, in the case of a single internal boundary, we were unable to obtain physically consistent trajectories using the parameters shown in the graphs. Several other tests were conducted using different values, but even then, it was not possible to obtain a valid trajectory. The problem lies with the constraint on $\dot{\theta}$, as the values obtained were always imaginary, making it impossible to obtain a valid trajectory.

It is essential to note that we use the Schwarzschild spacetime with zero Lorentz violation ($l=0$) immersed in a uniform external magnetic field to define each of the energy boundary cases. On this basis, we can make some detailed observations regarding the changes caused by the presence of the Lorentz violation parameter. In Fig. \ref{1caso}, which depicts the absence of an inner or outer boundary, we can see that when the values of l are included, the boundaries retain the same nature, with only minor changes in their extent. On the other hand, with regard to the trajectories, there are two escapes to infinity in the positive direction of the $z$-axis and a situation where the charged particle is absorbed by the black hole.

In Fig. \ref{2caso}, which shows the case of an inner boundary when $l$ is zero, we see a transition to both boundaries when $l$ is positive and the absence of boundaries when $l$ is negative. Furthermore, when there is an inner boundary, we obtain a trajectory that ends up being absorbed by the event horizon. When it changes to two boundaries, the result is a trajectory trapped around the horizon, and when there is no boundary, the trajectory obtained is also absorbed by the black hole.

Figure \ref{3caso} illustrates the case where both boundaries are present, in which the particle ends up being trapped in a region surrounding the event horizon. As can be seen, for $l=0$, the trajectory of the charged particle is confined to the region enclosed by the two boundaries. It turns out that when $l$ is positive, there is a sudden transition to an inner boundary, which consequently makes it impossible to obtain a physically acceptable trajectory. On the other hand, when $l$ is negative, we observe the absence of boundaries and a trajectory that descends and ascends, eventually being absorbed by the black hole.

Finally, in Fig. \ref{4caso}, which depicts the case of an inner boundary and the impossibility of obtaining a consistent physical trajectory, it can be seen that this remains true when the value of l is positive. However, when l becomes negative, the situation changes to that of an outer boundary, such that it is possible to obtain a trajectory that collapses into the black hole. We can therefore conclude that changes in the values of the Lorentz violation parameter cause significant, and sometimes extreme, changes in the boundaries generated by equation \eqref{E2}. Furthermore, small alterations in the other parameters, such as specific energy, specific angular momentum, etc., result in significant changes in the trajectories of the charged particles.

\begin{figure}[htpb]
\centering
\includegraphics[width=0.4\textwidth]{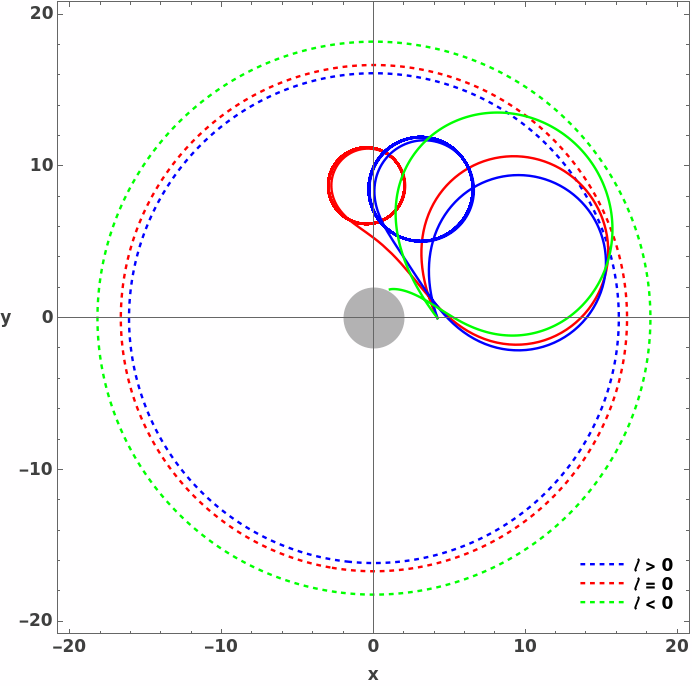}
\caption{A view in the $xy$ plane of the boundaries and trajectories of the first case, which is illustrated in Fig. \ref{1caso} with a view in the $xz$ plane.}
\label{1casoXY}
\end{figure}

The figures \ref{1caso}, \ref{2caso}, \ref{3caso}, \ref{4caso} illustrating the boundary cases show a view in the $xz$ plane. We will now examine the behaviour of these boundaries and trajectories in the $xy$ plane, as well as a three-dimensional view of some cases, as they are particularly interesting and warrant further comment. Figure \ref{1casoXY} shows the $xy$ cross-section of the first case. We can see how the outlines of the trajectories are similar, particularly the two that extend off into infinity.

\begin{figure}[htpb]
\centering
\includegraphics[width=0.4\textwidth]{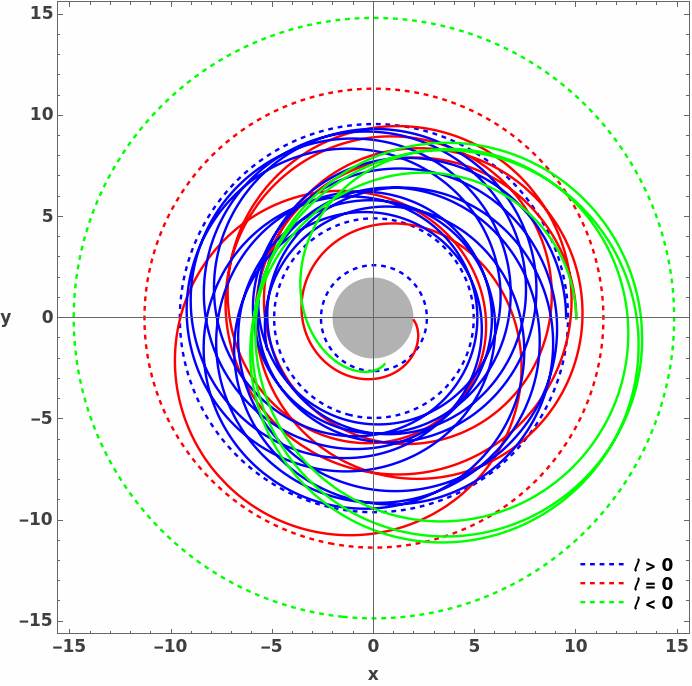}
\caption{View of the $xy$ plane for the second case, showing the boundaries and trajectories, as illustrated in Fig. \ref{2caso}.}
\label{2casoXY}
\end{figure}

In Fig. \ref{2casoXY}, which relates to the second case, we can clearly see the trajectories that are absorbed by the black hole, represented by the red and green curves. Furthermore, the trajectory that is trapped by the inner and outer boundaries is also clearly highlighted, as shown by the blue curve.

\begin{figure}[htpb]
\centering
\includegraphics[width=0.4\textwidth]{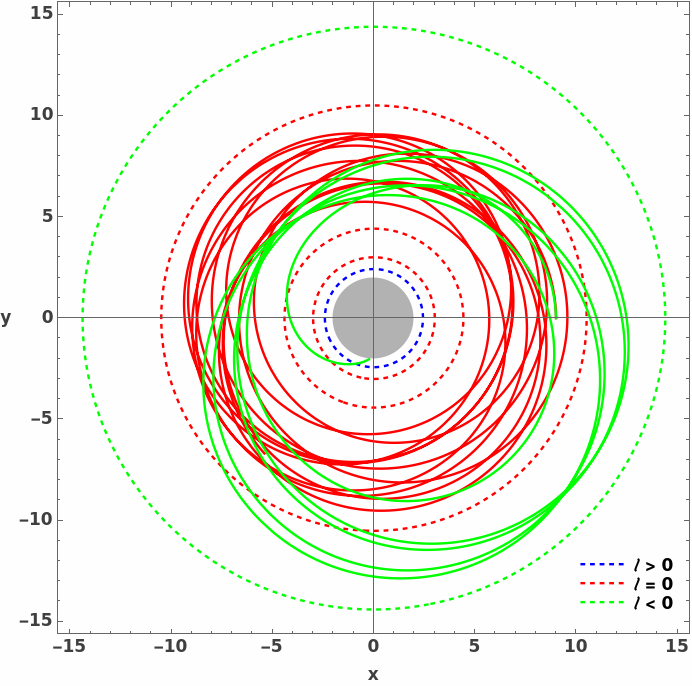}
\caption{Cross-section $xy$ showing the boundaries and trajectories for the third case, as shown in Fig. \ref{3caso}.}
\label{3casoXY}
\end{figure}

In the third case, we should bear in mind that for $l>0$ it was not possible to obtain a consistent trajectory, and for $l<0$ there is a transition to the boundary-less case with absorption of the charged particle. Figure \ref{3casoXY} provides a view of the $xy$ plane, allowing these changes in energy boundaries to be observed more clearly.

\begin{figure}[htpb]
\centering
\includegraphics[width=0.4\textwidth]{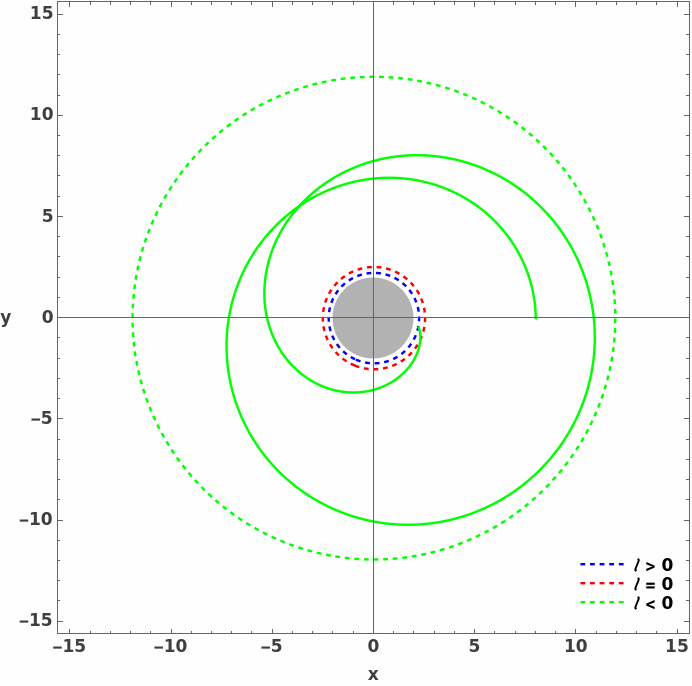}
\caption{$xy$ plane showing the boundaries and trajectories of the fourth case, as depicted in the $xz$ plane in Fig. \ref{4caso}.}
\label{4casoXY}
\end{figure}

Finally, in Fig. \ref{4casoXY} we have the $xy$ view of the fourth case. Once again, we see that the presence of an inner boundary is clearly established for $l=0$ and $l>0$. However, the transition to an outer boundary occurs when $l<0$, from which we can obtain an absorption trajectory.

\begin{figure}[htpb]
\centering
\includegraphics[width=0.4\textwidth]{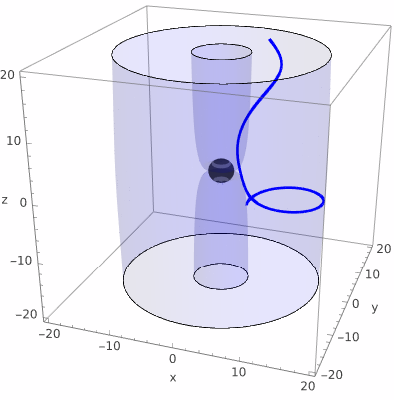}
\caption{Three-dimensional view of an escape trajectory corresponding to the first case. This trajectory is shown in Figures \ref{1caso} and \ref{1casoXY} for when $l>0$.}
\label{1caso3Dl+}
\end{figure}

\begin{figure}[htpb]
\centering
\includegraphics[width=0.4\textwidth]{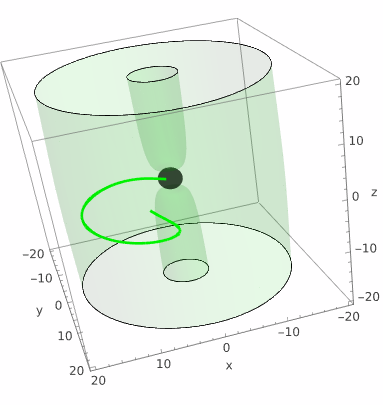}
\caption{Trajectory absorbed by the black hole, referring to the first case, when $l<0$. This trajectory is illustrated in Figs. \ref{1caso} and \ref{1casoXY}.}
\label{1caso3Dl-}
\end{figure}

We will now take a unique three-dimensional look at some of the trajectories shown in the previous figures. A comprehensive view provides a broader understanding of how energy boundaries are established and how the trajectories of charged particles are guided by these boundaries. Starting with the first case, there is no inner or outer boundary, so the particle can escape into infinity or even be swallowed by the black hole. Figure \ref{1caso3Dl+} illustrates an example of an escape trajectory, and Fig. \ref{1caso3Dl-} shows a situation where a particle is swallowed by the event horizon.

\begin{figure}[htpb]
\centering
\includegraphics[width=0.4\textwidth]{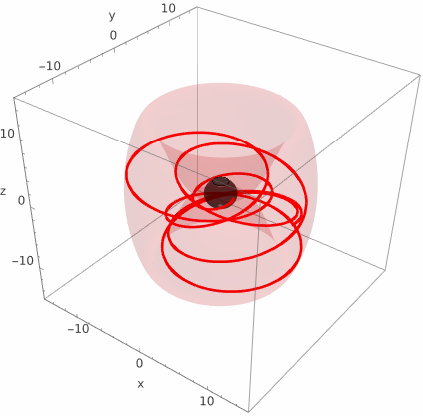}
\caption{Trajectory absorbed at an outer boundary, with $l=0$. This trajectory is also illustrated in Figs. \ref{2caso} and \ref{2casoXY}.}
\label{2caso3Dl0}
\end{figure}

\begin{figure}[htpb]
\centering
\includegraphics[width=0.4\textwidth]{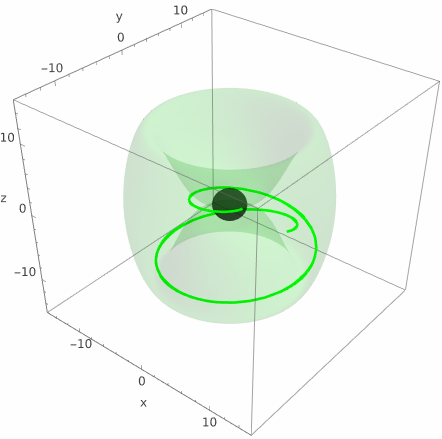}
\caption{A trajectory absorbed in a situation involving an outer boundary, where $l<0$. This trajectory is illustrated in Figures \ref{4caso} and \ref{4casoXY}.}
\label{4caso3Dl-}
\end{figure}

Figure \ref{2caso3Dl0} shows an absorbed trajectory for an outer boundary case (second case), with $l=0$. As can be seen, the particle even attempts to follow a stable trajectory around the event horizon, but is eventually absorbed after a certain period of time. On the other hand, this type of boundary also appears in a different form, in a case where $l<0$. We can see this in Fig. \ref{4caso3Dl-}.

\begin{figure}[htpb]
\centering
\includegraphics[width=0.4\textwidth]{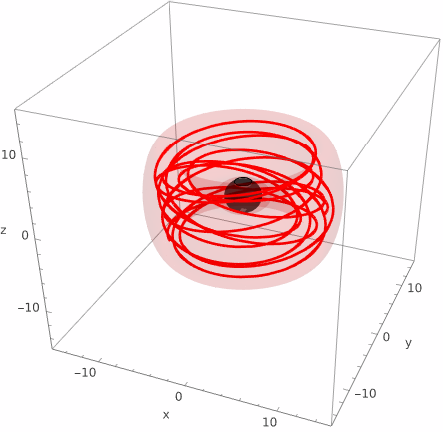}
\caption{Three-dimensional view of a trapped trajectory around the black hole, corresponding to the third case, with $l=0$. The other representations of this trajectory are shown in Figs. \ref{3caso} and \ref{3casoXY}.}
\label{3caso3Dl0}
\end{figure}

\begin{figure}[htpb]
\centering
\includegraphics[width=0.39\textwidth]{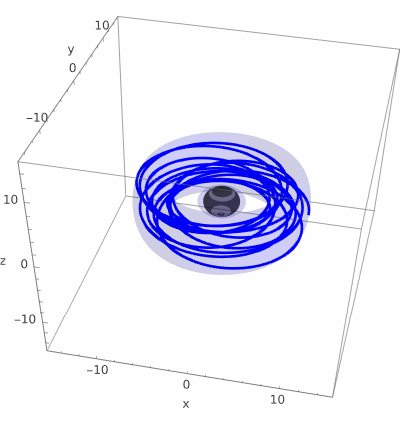}
\caption{A bound trajectory around the black hole, in the presence of both energy boundaries. In this case, we have $l>0$. The other representations of this trajectory are shown in Figs. \ref{2caso} and \ref{2casoXY}.}
\label{2caso3Dl+}
\end{figure}

The three-dimensional view of the inner and outer boundaries is shown in Fig. \ref{3caso3Dl0}. In this illustration, there is no violation of the Lorentz condition, $l=0$. We can also observe that the trajectory of the charged particle is partially stable and remains confined within a toroidal region around the event horizon. This type of trajectory with both boundaries also appears in another configuration when $l>0$, as can be seen in Fig. \ref{2caso3Dl+}.

\subsection{Curled trajectories}

So far, we have studied various types of trajectories, such as those escaping to infinity, those absorbed by the black hole, or even those confined to a region around the event horizon. All these types of trajectories are important and help us to understand the behaviour of charged particles, as well as the regions accessible for motion in accordance with the energy boundaries. However, amongst these trajectories, particularly those confined to a single region, there may be a very specific type of curled motion. This specific type of motion often follows a certain pattern, or comes very close to one. Consequently, the trajectory of the charged particle can be approximated as a periodic or quasi-periodic harmonic motion.

As mentioned earlier, there are certain conditions that must be met in order to obtain curled trajectories. One of these conditions concerns the values of the parameters $\mathcal{E}$ and $\mathcal{L}$; these values are determined by the functions $\mathcal{E}_{*}(\mathcal{L})$ and $\mathcal{L}_{*}(r)$ defined in equations \eqref{E*} and \eqref{L*}, which depend on the specific angular momentum $\mathcal{L}$ and the radial coordinate $r$, respectively. Another condition is that the magnetic field parameter $\mathcal{B}$ must be strictly positive. Thus, by referring to the graphs in Figures \ref{Lr} and \ref{EL}, for cases where $\mathcal{B}>0$, it was possible to obtain a number of curled trajectories for various values of $l$, in order to compare the changes caused by the presence of the Lorentz violation parameter. Furthermore, according to the graphs of the energy and angular momentum functions, very specific parameter values are required for these curled trajectories to occur.

\begin{figure*}[t]
	\centering
	\subfigure[\label{Curled01XZl0}]{\includegraphics[width=0.32\textwidth]{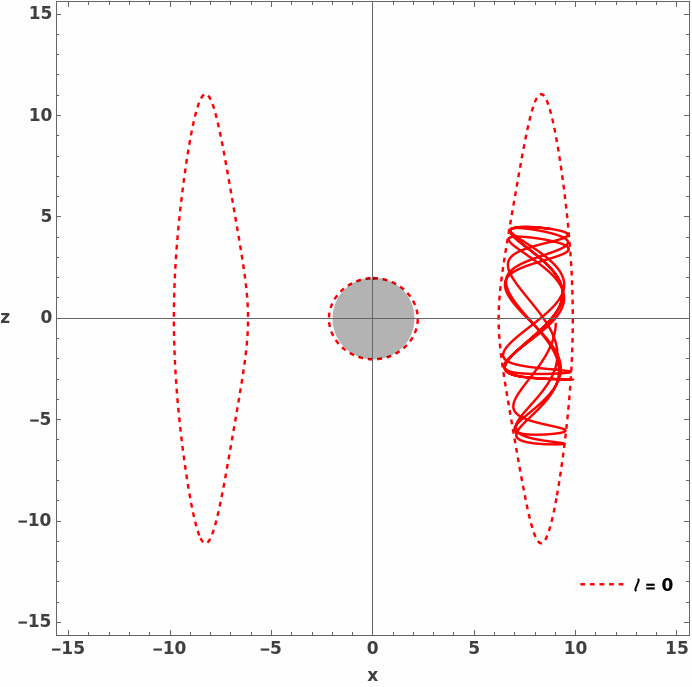}}
	\hfill
	\subfigure[\label{Curled01XZl+}]{\includegraphics[width=0.32\textwidth]{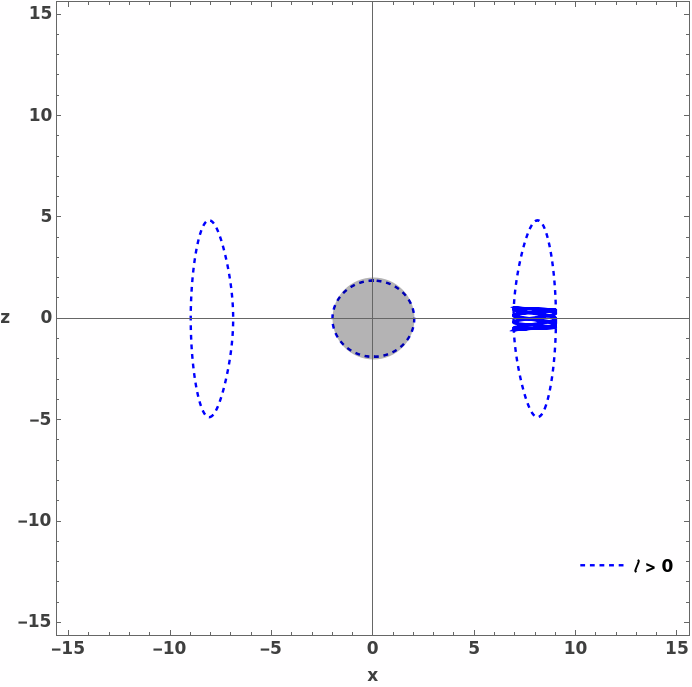}}
	\hfill
	\subfigure[\label{Curled01XZl-}]{\includegraphics[width=0.32\textwidth]{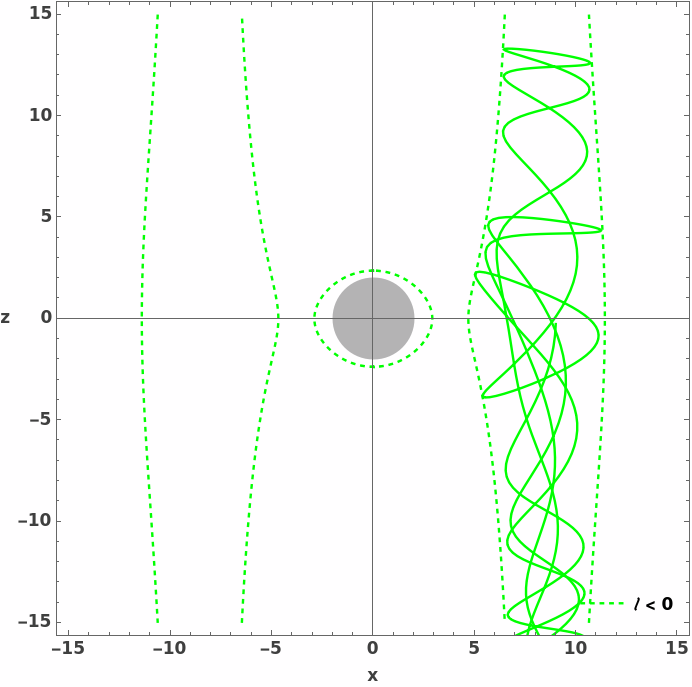}}
	\hfill
	\subfigure[\label{Curled01XYl0}]{\includegraphics[width=0.32\textwidth]{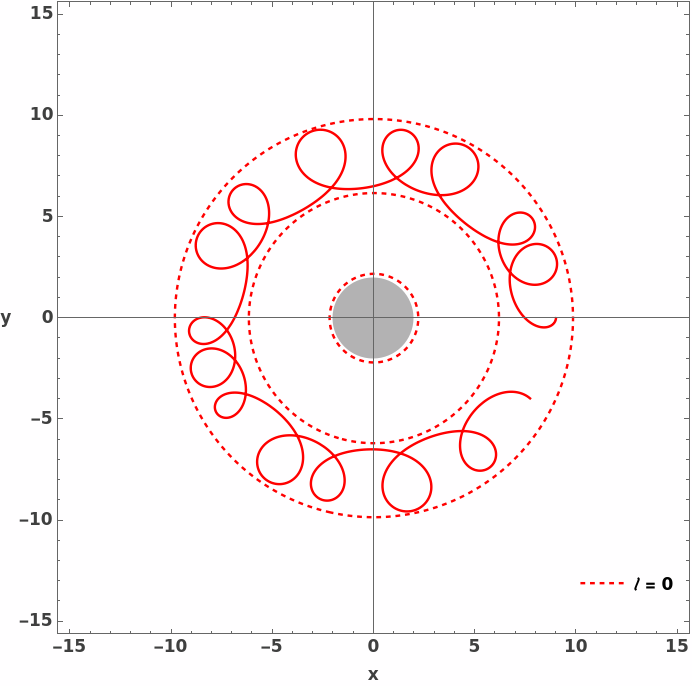}}
	\hfill
	\subfigure[\label{Curled01XYl+}]{\includegraphics[width=0.32\textwidth]{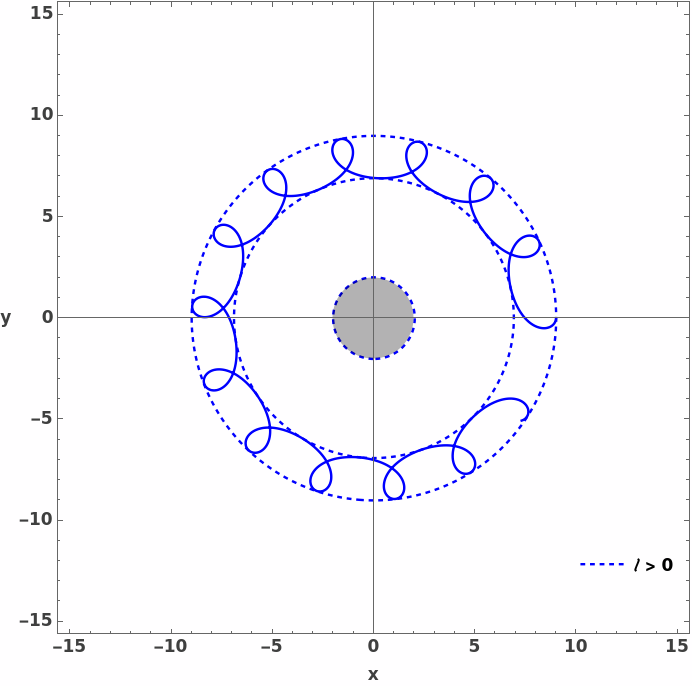}}
	\hfill
	\subfigure[\label{Curled01XYl-}]{\includegraphics[width=0.32\textwidth]{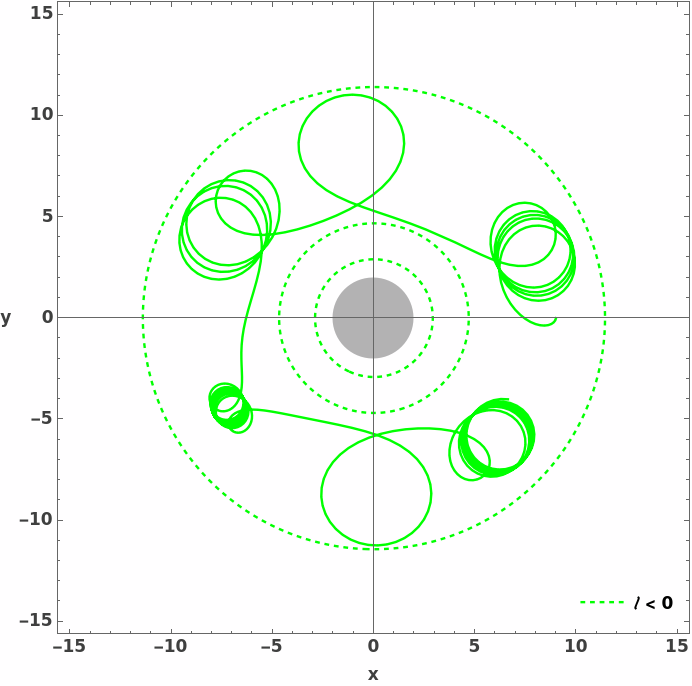}}
	\hfill
	\subfigure[\label{Curled013Dl0}]{\includegraphics[width=0.32\textwidth]{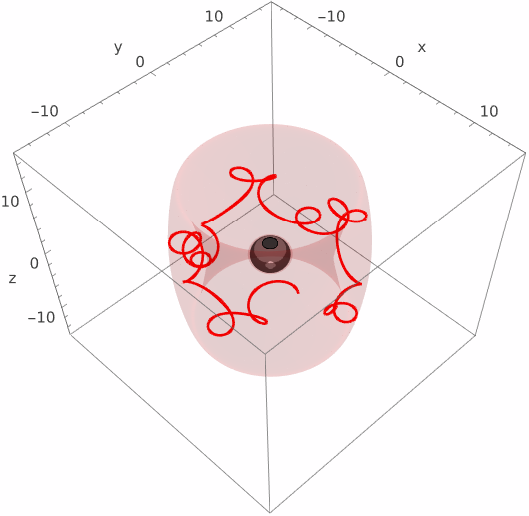}}
	\hfill
	\subfigure[\label{Curled013Dl+}]{\includegraphics[width=0.32\textwidth]{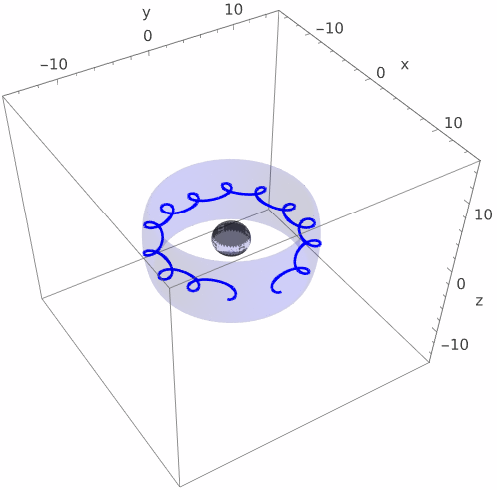}}
	\hfill
	\subfigure[\label{Curled013Dl-}]{\includegraphics[width=0.32\textwidth]{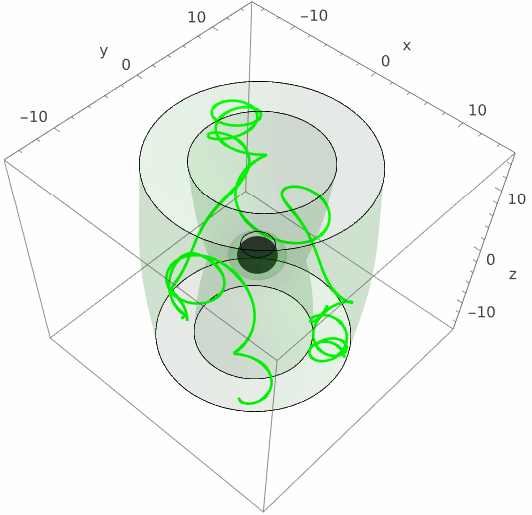}}
	\caption{Curled trajectories in the $xz$ and $xy$ planes and in three dimensions. For these trajectories, the parameters $r_{0}=9$, ${\theta}_{0}=1.6$, $\mathcal{L}=7$, $\mathcal{B}=0.1$, $\mathcal{E}=0.925$ and $l=(-0.185022, 0.060938)$ were used.}
	\label{Curled01}
\end{figure*}

We then selected the appropriate parameters and ran simulations to obtain the curled trajectories, along with projections onto the $xz$ and $xy$ planes, as well as a three-dimensional view. In addition to the specific parameters governing the trajectories, we carried out tests for various values of $l$ in order to compare the behaviour of charged particles with and without Lorentz violation. Figure \ref{Curled01} shows the trajectories projected onto the $xz$ and $xy$ planes, along with the three-dimensional view, and the corresponding energy surfaces.

From the $xy$ plane view, we can clearly see the curves formed along the trajectory and how they follow a certain pattern of motion, particularly in Fig. \ref{Curled01XYl+}. This pattern of motion characterises periodic, or quasi-periodic, motion. On the other hand, from the $xz$ plane, it is possible to observe the extent to which the trajectory is perturbed and deviates from the equatorial plane, due to the magnetic field and also to the Lorentz violation parameter. The three-dimensional view gives us an overview of the behaviour of the trajectories and how far they deviate from the equatorial plane. Fig. \ref{Curled013Dl+} shows a fairly stable trajectory, whilst Fig. \ref{Curled013Dl-} shows a highly perturbed trajectory.

\section{Frequencies of quasi-periodic oscillations}\label{sec4}

As we saw in the previous section, curled trajectories follow a pattern that may be periodic or approximately periodic, so they can be treated as harmonic motion. From this, we can determine the frequency values and make comparisons in order to understand the behaviour of the object and the characteristics of the astrophysical object that is guiding the movement of the charged particle. We will therefore examine the fundamental frequencies associated with the motion of particles and then perform a statistical analysis of these frequencies in order to model astrophysical objects and determine their properties, such as mass and the innermost stable circular orbit.

The first frequency is the azimuthal frequency ${\Omega}_{\phi}$, also known as the Keplerian frequency. We can determine it using the following expression
\begin{equation}
	{\Omega}_{\phi}^{2}=\frac{\dot{\phi}^{2}}{\dot{t}^{2}}.
	\label{OmegaK2}
\end{equation}

On the other hand, the most important frequencies for our study will be the radial frequency ${\Omega}_{r}$ and the angular frequency ${\Omega}_{\theta}$. These are the most important because there is a relationship between them, which will be essential for our model and statistical analysis. To determine the expressions for these frequencies, let us start by introducing small perturbations along these directions around the circular orbit
\begin{equation}
	r\rightarrow r_{0}+{\delta}r,
	\label{deltar}
\end{equation}
\begin{equation}
	{\theta}\rightarrow{\theta}_{0}+{\delta}{\theta}.
	\label{deltatheta}
\end{equation}

We can therefore rewrite the effective potential as a series of powers in terms of $r$ and $\theta$. Thus
\begin{equation}
	\begin{split}
		V_{\mathrm{eff}}(r,{\theta})=V_{\mathrm{eff}}(r_{0},{\theta}_{0})+\left.{\delta}r{\partial}_{r}V_{\mathrm{eff}}(r,{\theta})\right|_{r_{0},{\theta}_{0}} \\
		+\left.{\delta}{\theta}{\partial}_{\theta}V_{\mathrm{eff}}(r,{\theta})\right|_{r_{0},{\theta}_{0}}+\left.\frac{1}{2}{\delta}r^{2}{\partial}^{2}_{r}V_{\mathrm{eff}}(r,{\theta})\right|_{r_{0},{\theta}_{0}} \\
		+\left.\frac{1}{2}{\delta}{\theta}^{2}{\partial}^{2}_{\theta}V_{\mathrm{eff}}(r,{\theta})\right|_{r_{0},{\theta}_{0}}+\left.{\delta}r{\delta}{\theta}{\partial}_{r}{\partial}_{\theta}V_{\mathrm{eff}}(r,{\theta})\right|_{r_{0},{\theta}_{0}} \\
		+\mathcal{O}\left({\delta}r^{3},{\delta}{\theta}^{3}\right).
	\end{split}
	\label{Veffserie}
\end{equation}

The first term of the expansion vanishes due to the condition ${\partial}_{r}V_{\mathrm{eff}}=0$. The second term is eliminated due to the stability condition, which was expressed in equation \eqref{drdtheta}. These conditions also help to eliminate the remaining terms of the series, leaving only the second-order terms, which will be of interest to us when modelling harmonic oscillations. Thus, substituting what remains of the expansion into the effective potential, taking the equatorial plane, and rewriting the expressions in terms of coordinated time, we obtain the equations for the harmonic oscillations for the perturbations ${\delta}r$ and ${\delta}{\theta}$ as
\begin{equation}
	\frac{d^{2}{\delta}r}{dt^{2}}+{\Omega}^{2}_{r}{\delta}r=0,
	\label{deltar2}
\end{equation}
\begin{equation}
	\frac{d^{2}{\delta}{\theta}}{dt^{2}}+{\Omega}^{2}_{\theta}{\delta}{\theta}=0.
	\label{deltatheta2}
\end{equation}

The terms ${\Omega}^{2}_{r}$ and ${\Omega}^{2}_{\theta}$ are the radial and angular frequencies, respectively, as measured by a distant observer. These frequencies are defined by
\begin{equation}
	{\Omega}^{2}_{r}=\left.-\frac{1}{2g_{rr}\dot{t}^{2}}{\partial}^{2}_{r}V_{\mathrm{eff}}(r,{\theta})\right|_{{\theta}={\pi}/2},
	\label{Omega2r}
\end{equation}
\begin{equation}
	{\Omega}^{2}_{\theta}=\left.-\frac{1}{2g_{{\theta}{\theta}}\dot{t}^{2}}{\partial}^{2}_{\theta}V_{\mathrm{eff}}(r,{\theta})\right|_{{\theta}={\pi}/2}.
	\label{Omega2theta}
\end{equation}

Explicitly, using the effective potential~\eqref{Veff} and evaluating the second derivatives at the equatorial plane, the radial and vertical epicyclic frequencies are given by
\begin{equation}
	{\Omega}^{2}_{r}=\left.-\frac{f(r)}{2\dot{t}^{2}}{\partial}^{2}_{r}V_{\mathrm{eff}}(r,{\theta})\right|_{{\theta}={\pi}/2},
	\label{Omega2r_explicit}
\end{equation}
\begin{equation}
	{\Omega}^{2}_{\theta}=\left.-\frac{1}{2r^{2}\dot{t}^{2}}{\partial}^{2}_{\theta}V_{\mathrm{eff}}(r,{\theta})\right|_{{\theta}={\pi}/2},
	\label{Omega2theta_explicit}
\end{equation}
where ${\partial}^{2}_{\theta}V_{\mathrm{eff}}|_{{\theta}={\pi}/2}=2f(r)\bigl(\frac{\mathcal{L}^{2}}{r^{2}}-\mathcal{B}^{2}r^{2}\bigr)$, showing that the vertical epicyclic frequency depends explicitly on the magnetic field parameter $\mathcal{B}$.

For the purposes of a more detailed study, we can express the frequencies in SI units by converting
\begin{equation}
	{\nu}=\frac{1}{2{\pi}}\frac{c^{3}}{GM}{\Omega}.
	\label{nu}
\end{equation}

Thus, we performed some tests for the QPO frequencies ${\omega}_{\phi}$, ${\omega}_{\theta}$, and ${\omega}_{r}$. First, we considered the case without the Lorentz violation parameter $l$, with only a change in the magnetic field strength, and also compared it to the pure Schwarzschild case. Next, we considered the presence of the parameter $l$ associated with various variations in the magnetic field values. The results of these tests for the QPO frequencies are shown in the following figures. Figure \ref{QPOl0} shows the case where $l=0$, for the magnetic field parameter $\mathcal{B}$ when it is zero, positive, and negative.

\begin{figure*}[t]
	\centering
	\subfigure[\label{QPOB0l0}]{\includegraphics[width=0.32\textwidth]{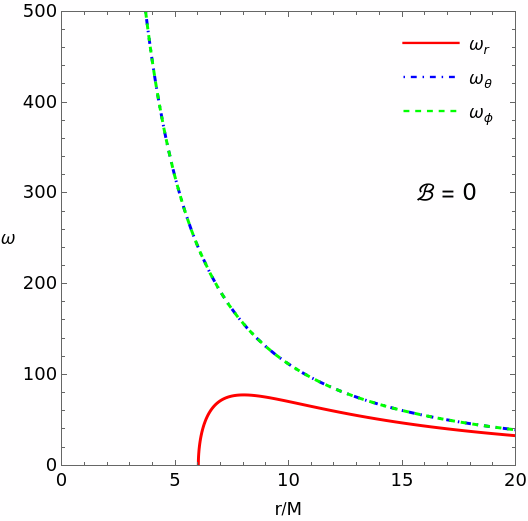}}
	\hfill
	\subfigure[\label{QPOB+l0}]{\includegraphics[width=0.32\textwidth]{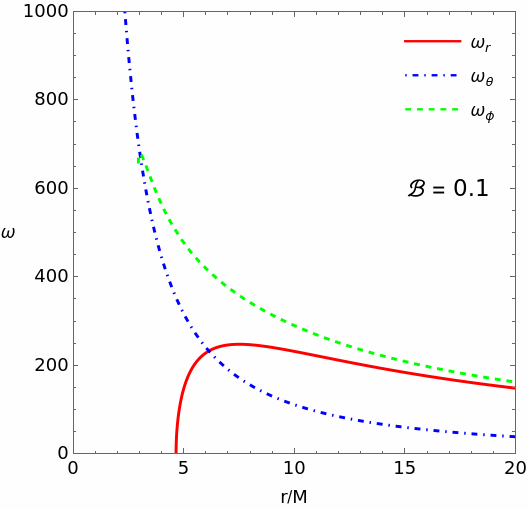}}
	\hfill
	\subfigure[\label{QPOB-l0}]{\includegraphics[width=0.32\textwidth]{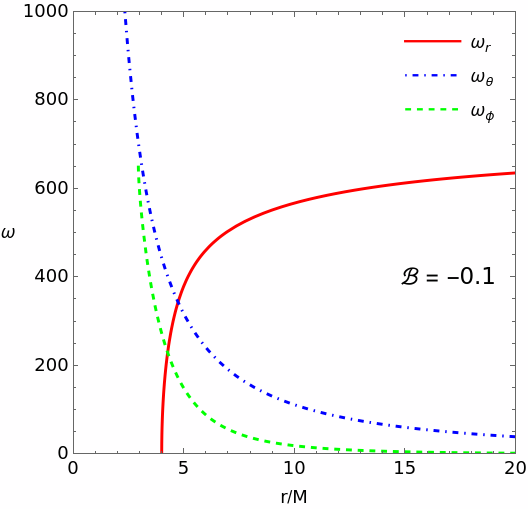}}
	\caption{Frequencies ${\omega}_{\phi}$, ${\omega}_{\theta}$, and ${\omega}_{r}$ for the case without Lorentz violation, with the magnetic field varying between zero, positive, and negative values.}
	\label{QPOl0}
\end{figure*}

As expected, the frequencies ${\omega}_{\phi}$ and ${\omega}_{\theta}$ are equal when there is no magnetic field. However, when we introduce the magnetic field, ${\omega}_{\phi}$ differs from ${\omega}_{\theta}$, which undergoes a small shift due to the $\mathcal{B}^{2}$ dependence in the second derivative of the effective potential. On the other hand, the frequency ${\omega}_{r}$ undergoes a very abrupt change in its magnitude when we alter the intensity of the parameter $\mathcal{B}$.

\begin{figure*}[t]
	\centering
	\subfigure[\label{omegaphiB0}]{\includegraphics[width=0.32\textwidth]{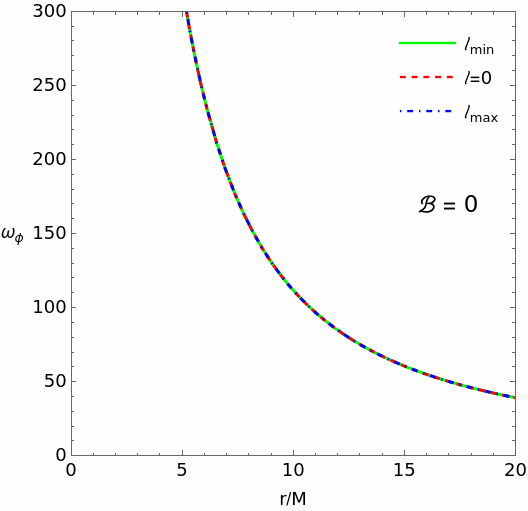}}
	\hfill
	\subfigure[\label{omegathetaB0}]{\includegraphics[width=0.32\textwidth]{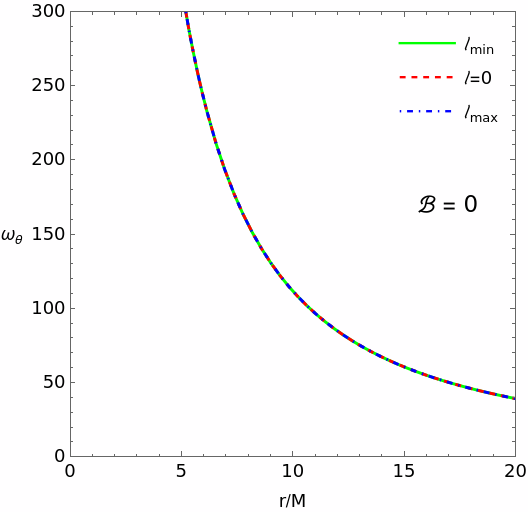}}
	\hfill
	\subfigure[\label{omegarB0}]{\includegraphics[width=0.32\textwidth]{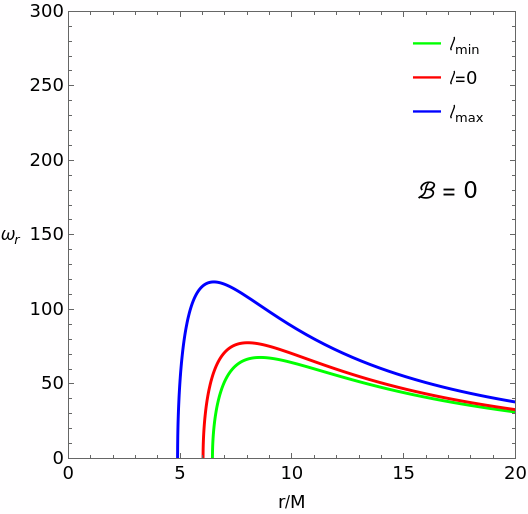}}
	\caption{Frequencies ${\omega}_{\phi}$, ${\omega}_{\theta}$, and ${\omega}_{r}$ in the absence of a magnetic field and with variations in the parameter $l$.}
	\label{omegasB0}
\end{figure*}

Now we will see how the frequencies behave in the presence of the Lorentz violation parameter associated with the external magnetic field. Following~\cite{Ednaldo2024}, we adopt a representative range for the Lorentz violation parameter: a maximum value of $l=0.189785$ and a minimum value of $l=-0.700225$. In the first case, we have the frequencies for the parameter $l$, without the magnetic field. The results are shown in Fig. \ref{omegasB0}, with particular attention given to the pure Schwarzschild case (red curves), which we can compare with the frequencies in Fig. \ref{QPOB0l0}.

In this case, when only the Lorentz violation parameter is considered, there are no changes in the frequencies ${\omega}_{\phi}$ and ${\omega}_{\theta}$, which remain degenerate, as expected for a spherically symmetric spacetime without a magnetic field. On the other hand, there are changes in the magnitude of ${\omega}_{r}$.

\begin{figure*}[t]
	\centering
	\subfigure[\label{omegaphiB+}]{\includegraphics[width=0.32\textwidth]{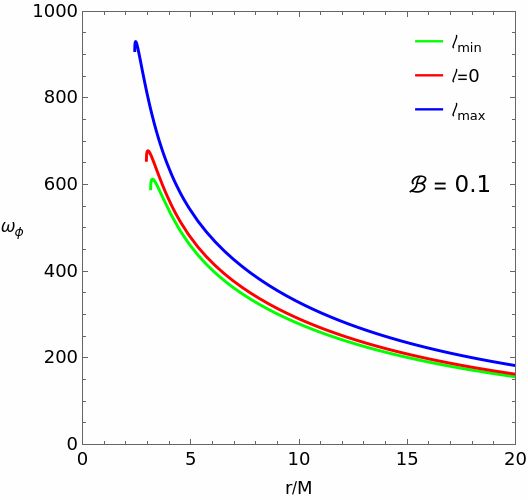}}
	\hfill
	\subfigure[\label{omegathetaB+}]{\includegraphics[width=0.32\textwidth]{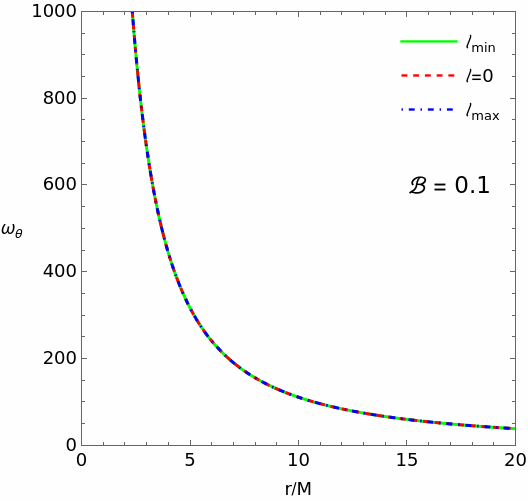}}
	\hfill
	\subfigure[\label{omegarB+}]{\includegraphics[width=0.32\textwidth]{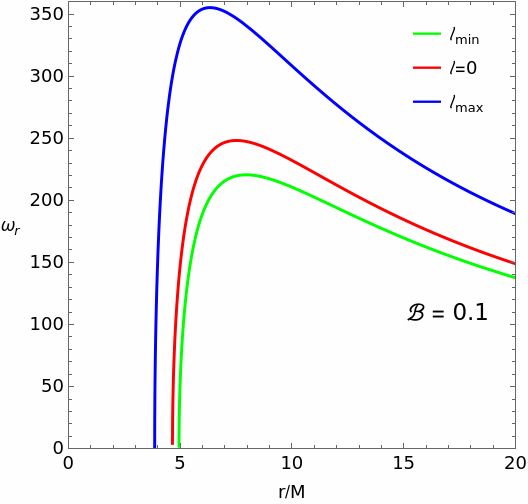}}
	\caption{Frequencies ${\omega}_{\phi}$, ${\omega}_{\theta}$, and ${\omega}_{r}$ with a positive magnetic field and variations in the parameter $l$.}
	\label{omegasB+}
\end{figure*}

Now, in Fig. \ref{omegasB+}, we have the three frequencies with the presence of a magnetic field of positive magnitude, associated with variations in the values of $l$. In this case, the red curves now represent the Schwarzschild case with only a magnetic field, as we saw in Fig. \ref{QPOB+l0}.

With the introduction of the magnetic parameter $\mathcal{B}$, we see that the frequency ${\omega}_{\phi}$ changes and varies in magnitude with $l$, while $\mathcal{B}$ remains constant. The frequency ${\omega}_{\theta}$ undergoes a small variation due to the combined effect of $\mathcal{B}$ and $l$, though this variation is less pronounced than that of ${\omega}_{\phi}$ or ${\omega}_{r}$ for the parameter range considered. Finally, the radial frequency undergoes a significant change in magnitude due to the inclusion of the magnetic field along with the Lorentz violation parameter.

\begin{figure*}[t]
	\centering
	\subfigure[\label{omegaphiB-}]{\includegraphics[width=0.32\textwidth]{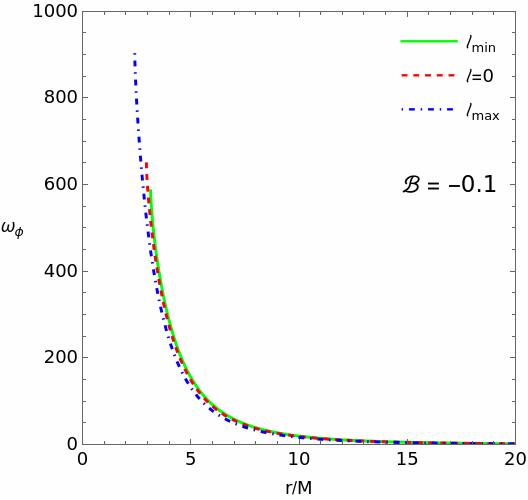}}
	\hfill
	\subfigure[\label{omegathetaB-}]{\includegraphics[width=0.32\textwidth]{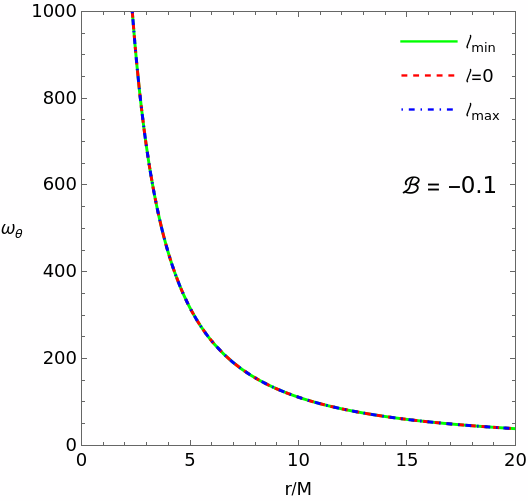}}
	\hfill
	\subfigure[\label{omegarB-}]{\includegraphics[width=0.32\textwidth]{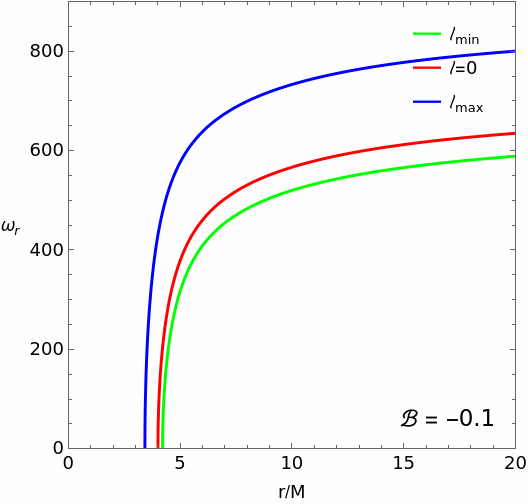}}
	\caption{Frequencies ${\omega}_{\phi}$, ${\omega}_{\theta}$, and ${\omega}_{r}$ with a negative magnetic field and variations in the parameter $l$.}
	\label{omegasB-}
\end{figure*}

Finally, we have the case with the negative magnetic field, associated with the Lorentz violation parameter. The frequencies obtained are shown in Fig. \ref{omegasB-}.

In this situation, we can observe that the frequency ${\omega}_{\phi}$ remains approximately stable and undergoes small variations in intensity. The frequency ${\omega}_{\theta}$ exhibits a modest dependence on $l$ through the metric function $f(r)$, while ${\omega}_{r}$ exhibits significant variations in intensity, especially when $l>0$. Again, we can compare the red curves to the Schwarzschild case with only a magnetic field, illustrated in Fig. \ref{QPOB-l0}.

\begin{figure}[htpb]
	\centering
	\includegraphics[width=0.4\textwidth]{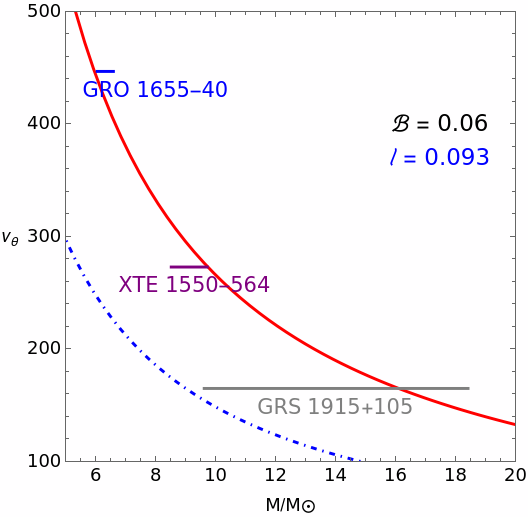}
	\caption{A model for the QPO frequencies of astrophysical objects with the associated parameters $\mathcal{B}$ and $l$.}
	\label{GRSB+l+}
\end{figure}

Regarding the modeling of QPO frequencies, we find that the inclusion of the magnetic field parameter is generally sufficient to break the degeneracy between the orbital and vertical epicyclic frequencies, allowing the fundamental frequencies to be fit to observational data. In the presence of a magnetic field, the Lorentz violation parameter $l$ provides an additional degree of freedom that can refine the model. In the absence of a magnetic field, the KR parameter alone affects the radial epicyclic frequency but cannot distinguish ${\omega}_{\phi}$ from ${\omega}_{\theta}$ due to the spherical symmetry of the spacetime. Nonetheless, when both $\mathcal{B}$ and $l$ are included, the combined effect allows for successful modeling of the observed QPO frequencies, as illustrated in Fig.~\ref{GRSB+l+}.

\begin{figure}[htpb]
	\centering
	\includegraphics[width=0.4\textwidth]{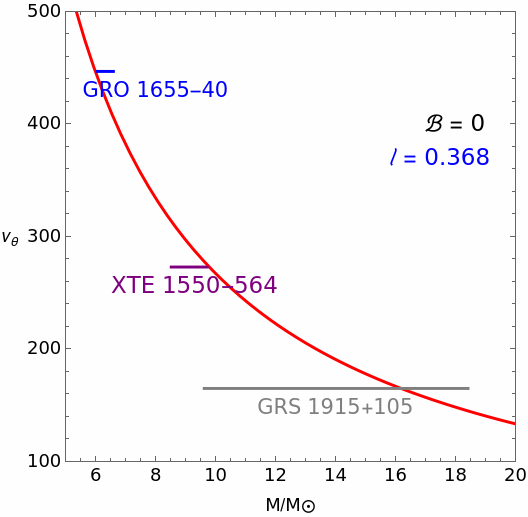}
	\caption{Modeling the QPO frequencies of astrophysical objects using the Lorentz violation parameter $l$ together with the magnetic field parameter $\mathcal{B}$.}
	\label{GRSB0l+}
\end{figure}

As shown in Fig.~\ref{GRSB+l+}, the value of $l$ decreases substantially due to the inclusion of the magnetic field parameter, which has a very low value. The red curve represents the $3:2$ resonance ratio, while the blue dashed-dotted curve represents the $2:3$ resonance ratio. Thus, the parameters $\mathcal{B}$ and $l$ work very well together in modeling QPO frequencies.

\subsection{Monte Carlo Markov Chain (MCMC) for quasi-periodic oscillations}

We will now use the Monte Carlo Markov chain statistical method to determine certain parameters of astrophysical objects based on the frequencies of quasi-periodic oscillations in the motion of charged particles \cite{Jumaniyozov2024,Guo2025,Cheng2023,Farukh2024,Ashraf2025,Nishonov2025,Ahal2026,Meng2026}. As a basis for the MCMC analysis, we will use three astrophysical objects that are candidates for microquasars: GRO 1655-40, XTE 1550-564, and GRS 1915+105. Based on the statistical analysis, we will determine the most appropriate values for the mass, the radius of the innermost stable circular orbit, magnetic field parameters, and Lorentz violation. Our goal in conducting these simulations is to verify the reliability of the Kalb-Ramond model in reproducing the frequencies of quasi-periodic oscillations observed in the astrophysical objects we are using as references in our research. We will perform an analysis without a magnetic field and Lorentz violation, and also in the presence of these parameters, in order to compare the main changes.

For our statistical analysis, we will use the \texttt{emcee} package from the Python library to implement the MCMC method. The results of the chain will be determined by the parameters of the astrophysical objects under study. The posterior distribution is expressed by the relation
\begin{equation}
	\mathcal{P}\left({\theta}|\mathcal{D},\mathcal{M}\right)=\frac{\mathcal{P}\left(\mathcal{D}|{\theta}\mathcal{M}\right){\pi}\left({\theta}|\mathcal{M}\right)}{\mathcal{P}\left(\mathcal{D}|\mathcal{M}\right)},
	\label{Pcali}
\end{equation}
where the function ${\pi}({\theta})$ is the prior function, and $\mathcal{P}\left(\mathcal{D}|{\theta}\mathcal{M}\right)$ is the likelihood function. For our simulations, we will consider Gaussian priors, such that ${\pi} ({\theta}_{i})\sim \exp{\left[\frac{1}{2}\left(\frac{{\theta}_{i}-{\theta}_{0,i}}{{\sigma}_{i}}\right)^{2}\right]}$, where ${\theta}_{\mathrm{low},i}<{\theta}_{i}<{\theta}_{\mathrm {high},i}$. In this case, the parameters involved are ${\theta}_{i}=\left\lbrace M, l, r/M \right\rbrace$ and ${\sigma}_{i}$ are their respective standard deviations.

Now, for a better and more comprehensive analysis, let us interpret the numerical data obtained in the following tables. This is necessary because graphs alone cannot provide us with a definitive answer as to which cases we can model the objects for the QPOs. To begin with, we have the likelihood function. This function is constructed from the independent Gaussian uncertainties for the observed frequencies. Thus, for a given set $\theta$, as defined above, the theoretical models for predicting the frequencies are compared to the observational data using a $\chi^2$ statistic. Therefore, the log-likelihood function can be expressed as
\begin{equation}
	\ln{\mathcal{L}\left({\theta}\right)}=-\frac{1}{2}\sum_{i}{\left(\frac{{\nu}_{i}^{\mathrm{obs}}({\theta})-{\nu}_{i}^{\mathrm{th}}({\theta})}{{\sigma}_{i}}\right)^{2}}.
	\label{loglikelihood}
\end{equation}

The Gaussian prior ensures that the values of the free parameters remain close to a central value. For the simulations, the Lorentz violation parameter $l$, as well as the magnetic field parameter $\mathcal{B}$, were restricted to the interval $\left[-0.5,0.5\right]$. In addition, we also restricted the value of the orbital radius, as follows
\begin{equation}
	\frac{r}{r_{ISCO}}>\frac{13}{12}.
	\label{rrisco}
\end{equation}

We added this constraint on the radius of the circular orbit for the following reason: for a pure Schwarzschild black hole, the radius of the innermost stable circular orbit is $r=6M$. For convenience, it is common to use orbits around $6.5M$ for simulations involving trajectories and, in our case, QPO frequencies. Since the frequencies were defined in terms of the dimensionless parameter $r/M=y$, where
\begin{equation}
	y=6\left(\frac{r}{r_{ISCO}}\right),
	\label{yrM}
\end{equation}
therefore, the constraint in equation \eqref{rrisco} allows the parameter to be at least $y=6.5$, keeping our study within the boundaries of the innermost stable circular orbit. This ensures that the orbit remains within the minimum limit allowed by the model we are using.

On the other hand, to determine whether the simulated model is suitable for our study with the objects we are analyzing, we must examine the values obtained for the AIC (Akaike Information Criterion) and BIC (Bayesian Information Criterion) parameters, as they will indicate the model's consistency and whether it is statistically viable. In MCMC analysis, the AIC parameter is defined as follows
\begin{equation}
	AIC=2k-\ln{\mathcal{L}_{\mathrm{max}}},
	\label{AIC}
\end{equation}
where $k$ represents the number of free parameters and $\mathcal{L}_{\mathrm{max}}$ is the maximum likelihood value obtained from the MCMC simulations. On the other hand, the BIC parameter is determined by the expression
\begin{equation}
	BIC=k\ln{N}-2\ln{\mathcal{L}_{\mathrm{max}}},
	\label{BIC}
\end{equation}
where $N$ is the number of observed data points, and $k$ and $\mathcal{L}_{\mathrm{max}}$ have already been defined above.

\begin{equation}
	{\Delta}AIC=AIC_{i}-AIC_{\mathrm{min}},
	\label{DeltaAIC}
\end{equation}
\begin{equation}
	{\Delta}BIC=BIC_{i}-BIC_{\mathrm{min}},
	\label{DeltaBIC}
\end{equation}
where $AIC_{i}$ and $BIC_{i}$ are the values of the models to be analyzed, and $AIC_{\mathrm{min}}$ and $BIC_{\mathrm{min}}$ are the minimum values obtained in the simulations, which will serve as the reference for the analysis.

Based on this, we have a reference scale for changes in AIC and BIC, which allows us to determine whether the model is statistically reliable. For changes in AIC, the smaller the change, the better the model; thus, the scale indicates the model's consistency. Therefore
\begin{itemize}
	\item $0\leq{\Delta}AIC\leq 2$: statistically equivalent models.
	\item $4\leq{\Delta}AIC\leq 7$: moderate evidence against the model.
	\item ${\Delta}AIC\geq 10$: strong evidence against the model.
\end{itemize}

On the other hand, changes in the BIC will indicate evidence against the model; thus, the smaller the change, the weaker the evidence and the more consistent the model.
\begin{itemize}
	\item $0\leq{\Delta}BIC<2$: weak evidence.
	\item $2\leq{\Delta}BIC<6$: positive evidence.
	\item $6\leq{\Delta}BIC<10$: strong evidence.
	\item ${\Delta}BIC\geq 10$: very strong evidence.
\end{itemize}

For our analysis, we will use the following ranges of values obtained from observational data

\begin{table}[h]
	\centering
	\begin{tabular}{|l|l|l|l|}
		\hline
		Source & GRO 1655-40 & XTE 1550-564 & GRS 1915+105 \\
		\hline
		${\nu}_{U}$ (Hz) & 447-453 & 273-279 & 165-171 \\
		${\nu}_{L}$ (Hz) & 295-305 & 179-189 & 108-118 \\
		$M/M_{\odot}$ & 6.03-6.57 & 8.5-9.7 & 9.6-18.4 \\
		\hline
	\end{tabular}
	\caption{QPO frequencies of twin peaks observed for three microquasars, along with their respective mass constraints. The terms ${\nu}_{U}$ and ${\nu}_{L}$ denote the upper and lower frequencies, respectively.}
	\label{microquasars}
\end{table}

Our analysis begins with $10$ chains of $100,000$ steps each for each parameter tested, resulting in $1,000,000$ samples of each parameter for each source. Before the posterior distribution converges, a portion of the samples is discarded; these are the burn-in values, estimated at $500,000$. Thus, each parameter of each source will have a total of $500,000$ valid samples in the statistical analysis. Using these reference data, an MCMC analysis was performed to determine the best values for mass, magnetic field parameters, Lorentz violation, and the radius of the innermost stable circular orbit. The measurements were carried out in accordance with Table \ref{microquasars}, with a total of $500,000$ samples for each simulation. Additionally, for comparison purposes, simulations were conducted for cases without a magnetic field and without the Lorentz violation parameter. In our simulations, we will consider three model configurations. The first will be a pure Schwarzschild case, where $l=0$ and $\mathcal{B}=0$. Next, we will consider the case with the inclusion of the Lorentz violation parameter, allowing it to take any value within the permitted range, while keeping the magnetic parameter zero, $\mathcal{B}=0$. Finally, the situation where the parameters $l$ and $\mathcal{B}$ are free to take any values appropriate for our research.

\begin{figure*}[t]
	\centering
	\subfigure[\label{GRO01}]{\includegraphics[width=0.32\textwidth]{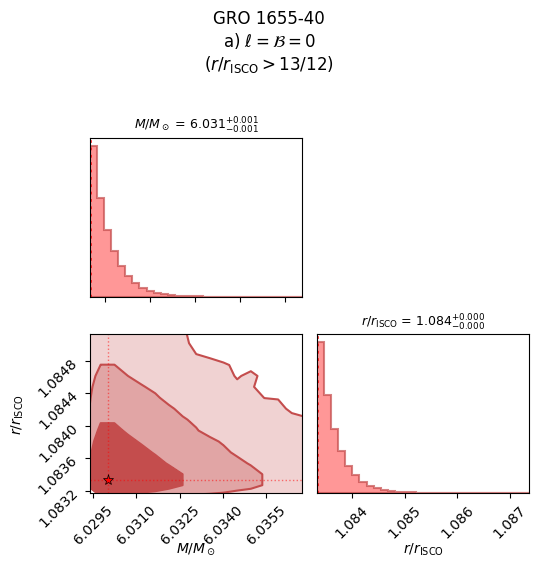}}
	\hfill
	\subfigure[\label{GRO02}]{\includegraphics[width=0.32\textwidth]{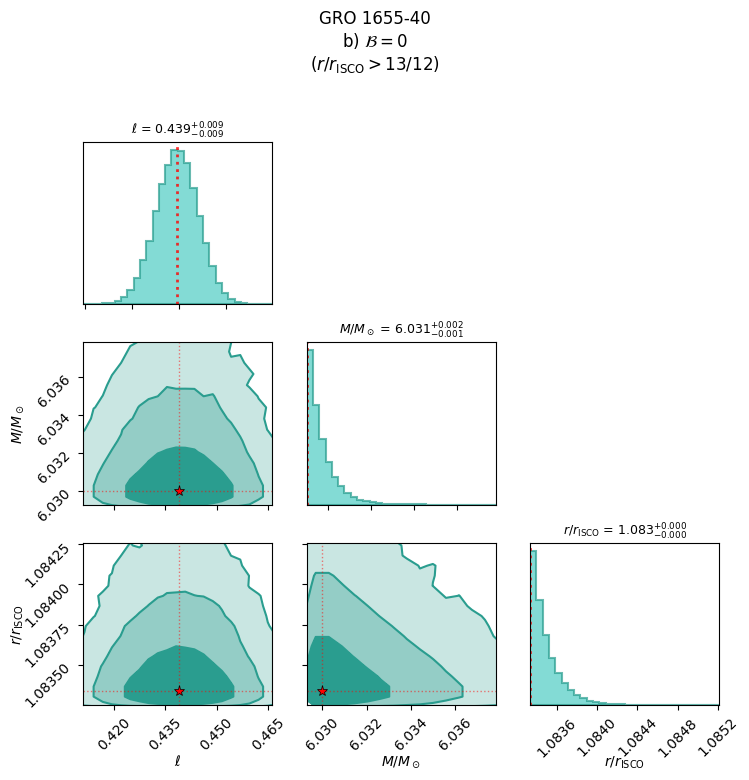}}
	\hfill
	\subfigure[\label{GRO03}]{\includegraphics[width=0.32\textwidth]{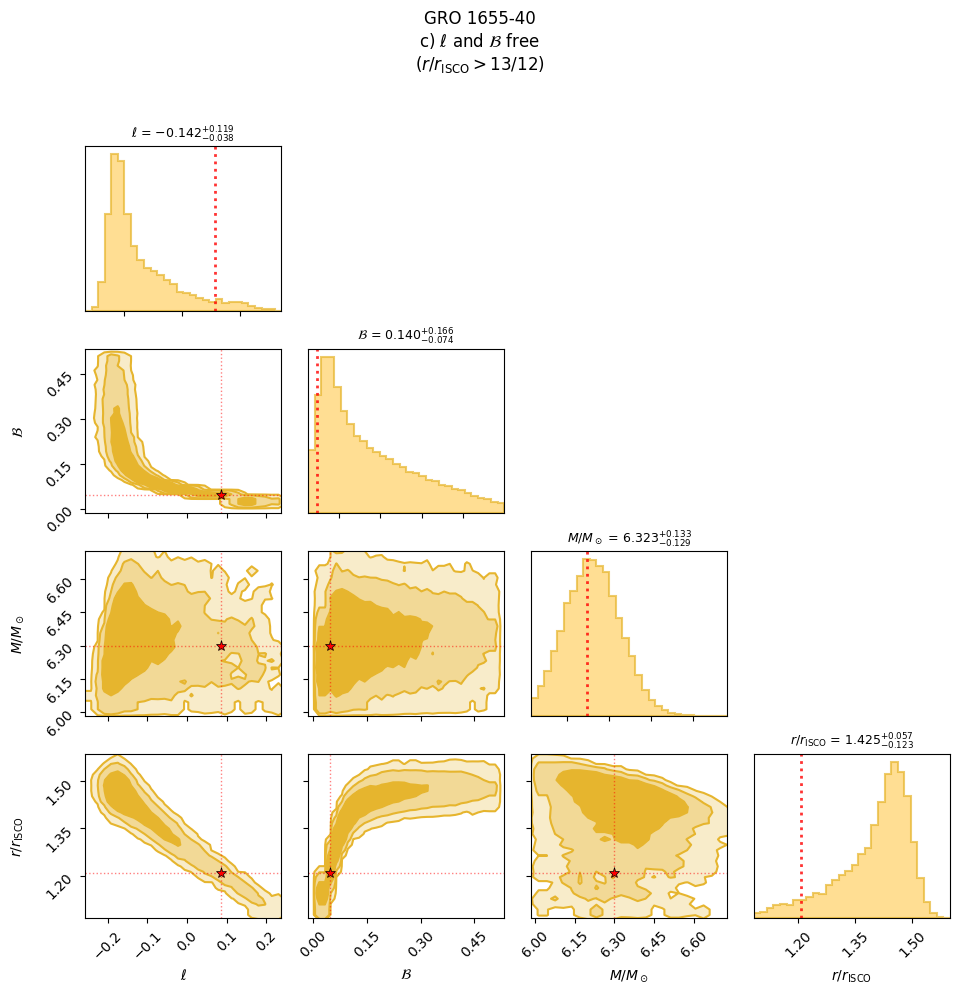}}
	\caption{Plots showing the confidence regions and distribution plots for the parameters of object GRO 1655-40. Scenarios were tested with and without a magnetic field and Lorentz violation, as well as in the presence of both. The red dashed lines (and the stars) indicate the best value for each quantity.}
	\label{GRO}
\end{figure*}

\begin{figure*}[t]
	\centering
	\subfigure[\label{XTE01}]{\includegraphics[width=0.32\textwidth]{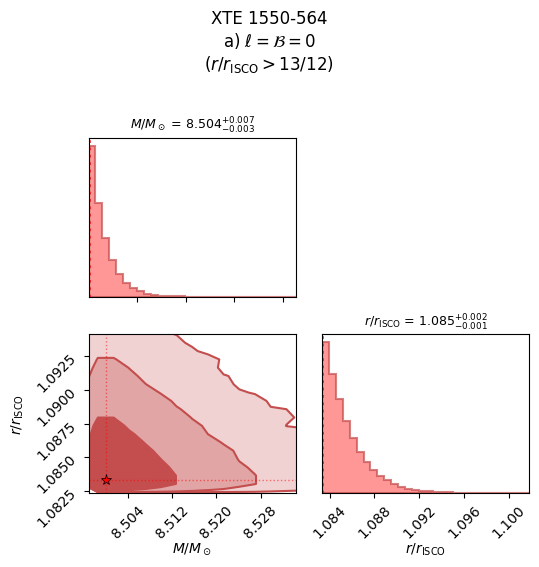}}
	\hfill
	\subfigure[\label{XTE02}]{\includegraphics[width=0.32\textwidth]{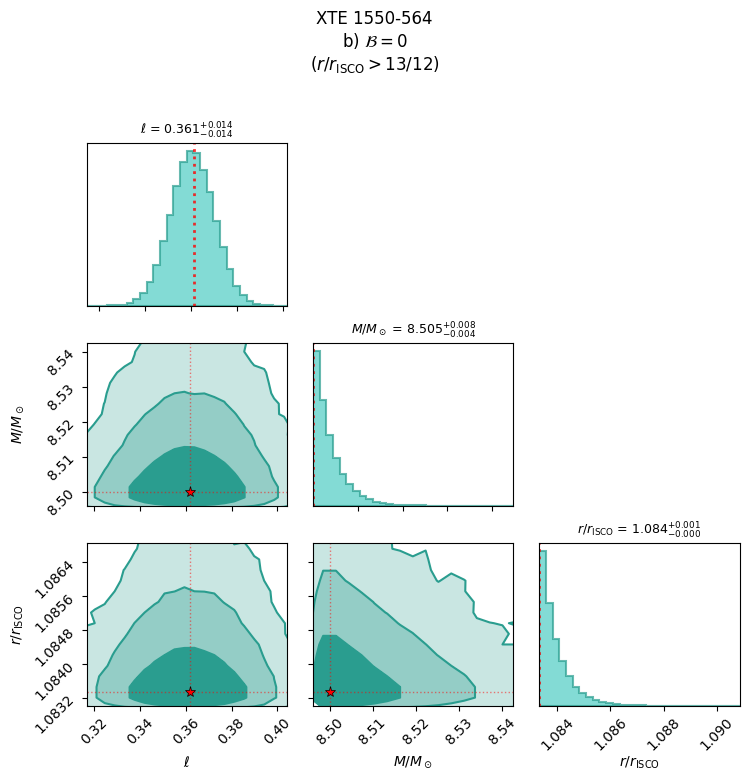}}
	\hfill
	\subfigure[\label{XTE03}]{\includegraphics[width=0.32\textwidth]{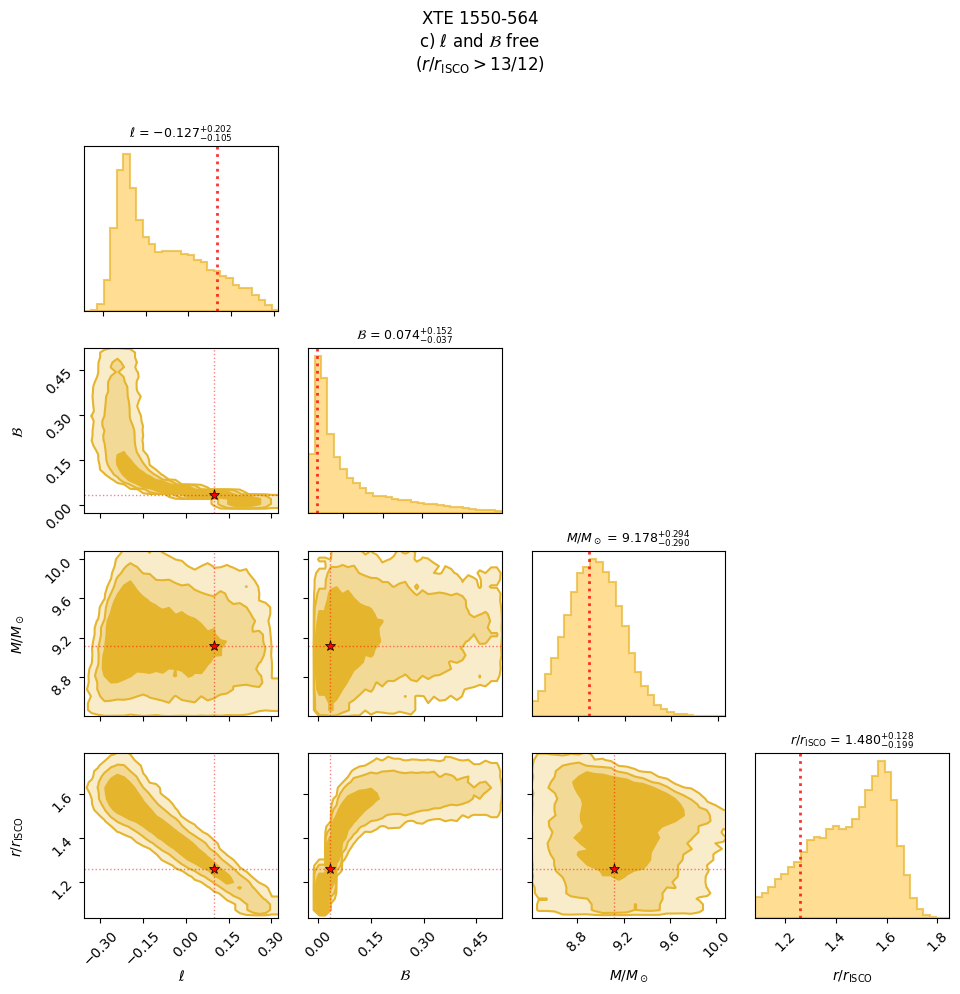}}
	\caption{Charts and distributions for object XTE 1550-564. The same scenarios and indicators mentioned in Fig. \eqref{GRO} were applied here.}
	\label{XTE}
\end{figure*}

\begin{figure*}[t]
	\centering
	\subfigure[\label{GRS01}]{\includegraphics[width=0.32\textwidth]{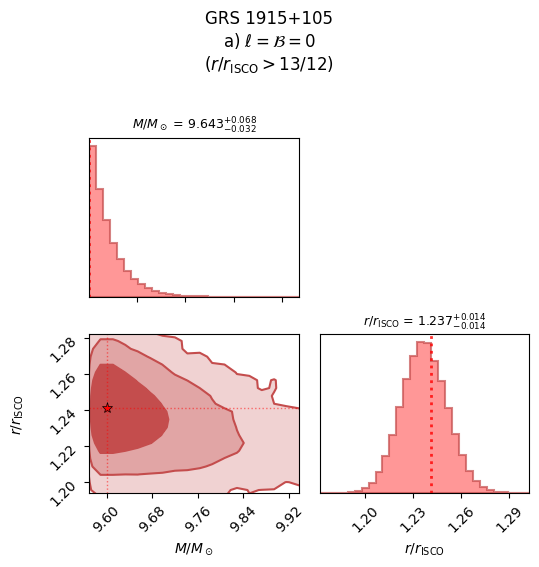}}
	\hfill
	\subfigure[\label{GRS02}]{\includegraphics[width=0.32\textwidth]{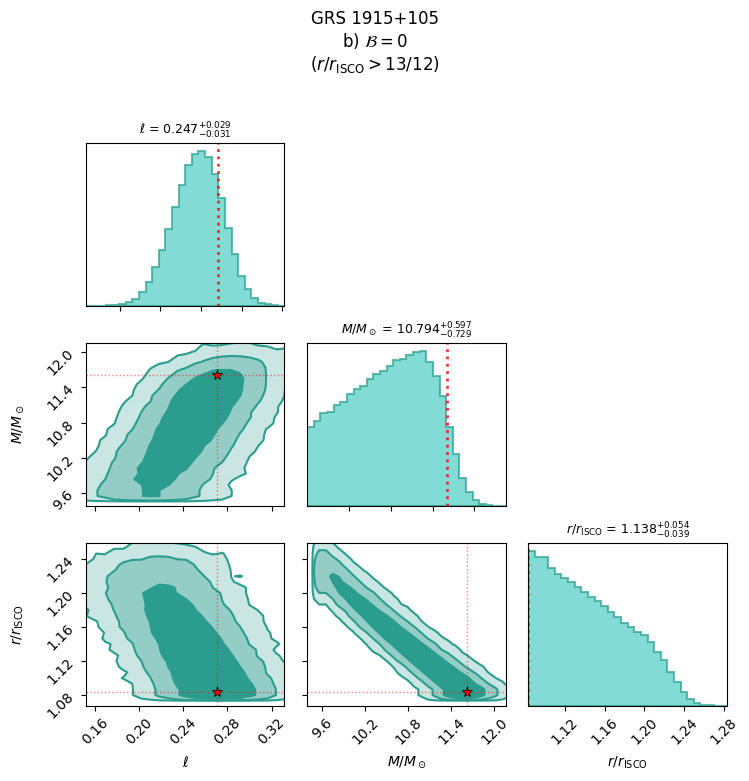}}
	\hfill
	\subfigure[\label{GRS03}]{\includegraphics[width=0.32\textwidth]{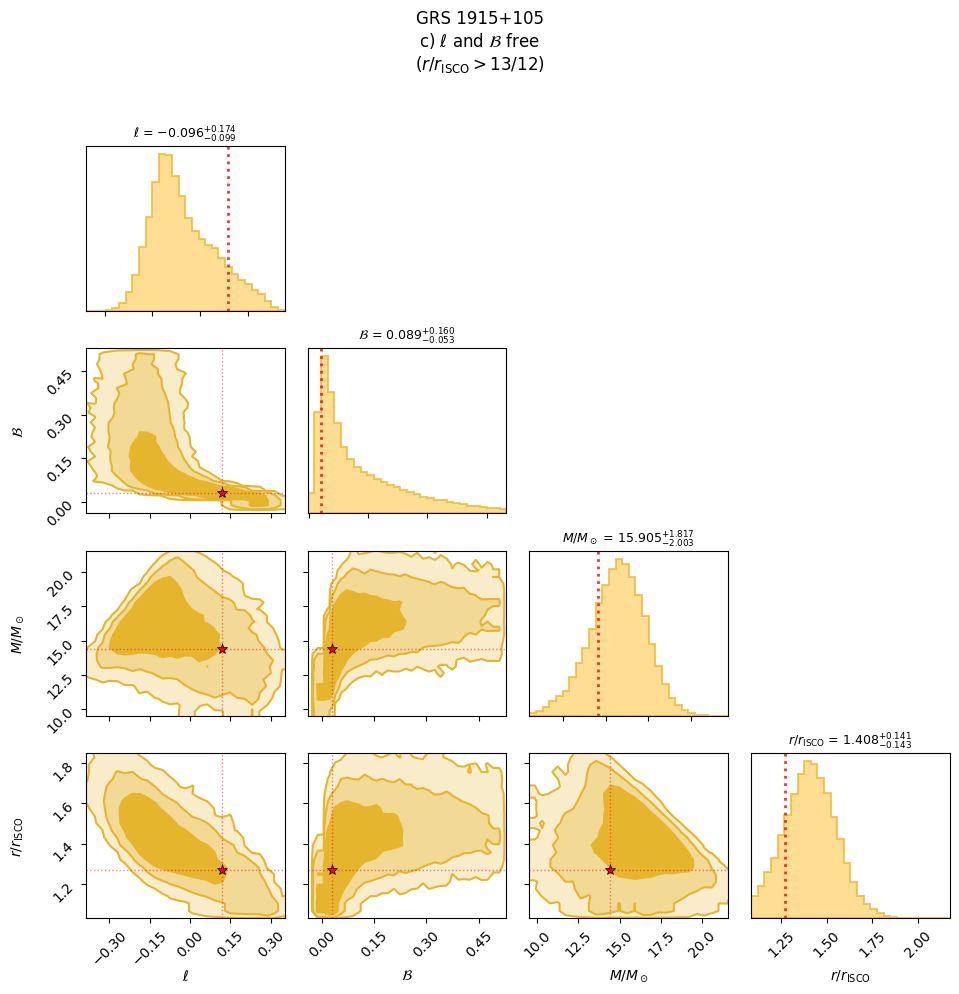}}
	\caption{Finally, the graphs and distributions for the object GRS 1915+105.}
	\label{GRS}
\end{figure*}

Figures \ref{GRO}, \ref{XTE}, and \ref{GRS} display the graphs for regions $1{\sigma}$, $2{\sigma}$, and $3{\sigma}$, respectively, from darkest to lightest, which represent the confidence interval for the quantities we are analysing, as well as the graphs illustrating the distribution of each quantity. It is important to emphasise that a Gaussian prior was used in these simulations.

In Fig. \ref{GRO}, we can see the results for the object GRO 1655-40. In this case, the measured parameters exhibit a positive trend, with the exception of a few notable points. In the case where $l$ and $\mathcal{B}$ are free, we observe that some of the best values are located within the $2{\sigma}$ region, but this is still within the confidence interval. This becomes more evident when we look at the distribution plots. However, we can note an improvement in the parameters, Fig. \ref{GRO03}, compared to the Schwarzschild case, Fig. \ref{GRO01}, since the best values are slightly further away from the limit imposed by the prior.

For the object XTE 1550-564, Fig. \ref{XTE}, we observed a similar improvement as we included the Lorentz violation parameter and the magnetic field. The shapes of the plots are somewhat similar to those obtained for the object GRO 1655-40. This demonstrates an improvement in the results obtained for this object as well. One point to highlight in the results for XTE 1550-564 is that all of its best values with Lorentz violation and magnetic field, Fig. \ref{XTE03}, fall within the $1{\sigma}$ confidence interval, confirming the accuracy of the measured values.

Finally, in the case of the object GRS 1915+105, shown in Fig. \ref{GRS}, we see that the best-fitting values also all lie within the $1{\sigma}$ region, indicating that the results obtained are quite accurate. On the other hand, the resulting plots are quite asymmetric, suggesting that, overall, the behavior deviates from the Gaussian distribution, which was the prior imposed in the simulations of these objects.

We used MCMC chains via the \texttt{emcee} package with $10$ independent chains and $100,000$ steps per chain. After this initial stage, we collected a total of $500,000$ samples for each case. From this, we report the best value, the mean (with the margin of error), the median, and the confidence intervals of the simulations: $68.27\%$ ($1{\sigma}$), $95.45\%$ ($2{\sigma}$), $99.73\%$ ($3{\sigma}$).

\begin{table*}[t]
	\centering
	\begin{tabular}{|l|l|l|l|l|l|l|}
		\hline
		Parameter & Best Value & Average $\pm{\sigma}$ & Median & $1{\sigma}$ & $2{\sigma}$ & $3{\sigma}$ \\
		\hline
		$l=\mathcal{B}=0$ & & & & & ${\Delta}AIC=3551.05$ & ${\Delta}BIC=3553.66$ \\
		\hline
		$M/M_{\odot}$ & $6.0300$ & $6.0312\pm 0.0012$ & $6.0308$ & $\left[6.0302,6.0322\right]$ & $\left[6.0300,6.0344\right]$ & $\left[6.0300,6.0375\right]$ \\
		$r/r_{ISCO}$ & $1.0833$ & $1.0836\pm 0.0003$ & $1.0835$ & $\left[1.0834,1.0839\right]$ & $\left[1.0833,1.0845\right]$ & $\left[1.0833,1.0854\right]$ \\
		Statistics & AIC: $3559.05$ & BIC: $3556.43$ & LogLik: $-1777.52$ & Nº Samples: $500,000$ & Nº Parameters: $2$ & $r_{min}=1.08333$ \\
		\hline
		$\mathcal{B}=0$ & & & & & ${\Delta}AIC=1783.21$ & ${\Delta}BIC=1784.52$ \\
		\hline
		$l$ & $0.4391$ & $0.4391\pm 0.0090$ & $0.4392$ & $\left[0.4302,0.4481\right]$ & $\left[0.4209,0.4568\right]$ & $\left[0.4115,0.4653\right]$ \\
		$M/M_{\odot}$ & $6.0300$ & $6.0314\pm 0.0013$ & $6.0309$ & $\left[6.0302,6.0325\right]$ & $\left[6.0300,6.0351\right]$ & $\left[6.0300,6.0389\right]$ \\
		$r/r_{ISCO}$ & $1.0833$ & $1.0835\pm 0.0002$ & $1.0834$ & $\left[1.0834,1.0836\right]$ & $\left[1.0833,1.0839\right]$ & $\left[1.0833,1.0844\right]$ \\
		Statistics & AIC: $1791.21$ & BIC: $1787.29$ & LogLik: $-892.61$ & Nº Samples $500,000$ & Nº Parameters: $3$ & $r_{min}=1.08333$ \\
		\hline
		$l$ and $\mathcal{B}$ free & & & & & ${\Delta}AIC=0.00$ & ${\Delta}BIC=0.00$ \\
		\hline
		$l$ & $0.0833$ & $-0.1075\pm 0.0907$ & $-0.1423$ & $\left[-0.1804,-0.0223\right]$ & $\left[-0.2041,0.1510\right]$ & $\left[-0.2235,0.2264\right]$ \\
		$\mathcal{B}$ & $0.0449$ & $0.1744\pm 0.1156$ & $0.1398$ & $\left[0.0651,0.3068\right]$ & $\left[0.0365,0.4498\right]$ & $\left[0.0276,0.4958\right]$ \\
		$M/M_{\odot}$ & $6.2970$ & $6.3257\pm 0.1280$ & $6.3227$ & $\left[6.1930,6.4568\right]$ & $\left[6.0842,6.5882\right]$ & $\left[6.0349,6.7204\right]$ \\
		$r/r_{ISCO}$ & $1.2073$ & $1.3984\pm 0.0946$ & $1.4254$ & $\left[1.3010,1.4823\right]$ & $\left[1.1500,1.5233\right]$ & $\left[1.0910,1.5577\right]$ \\
		Statistics & AIC: $8.00$ & BIC: $2.77$ & LogLik: $0.00$ & Nº Samples: $500,000$ & Nº parameters: $4$ & $r_{min}=1.08333$ \\
		\hline
	\end{tabular}
	\caption{Results obtained from the MCMC simulation for the object GRO 1655-40.}
	\label{GROdata1}
\end{table*}

Table \ref{GROdata1} presents the data for the object GRO 1655-40. As we can see, for the Schwarzschild case ($l=0$ and $\mathcal{B}=0$), the AIC and BIC values are very high, indicating that this model is not compatible with the QPOs. Furthermore, the ISCO value lies at the limit imposed by the Gaussian prior, and the object's mass is also close to the lower limit, confirming that this model is not suitable for our purposes. With the inclusion of the Lorentz violation parameter, we see a considerable improvement in the AIC and BIC values, but they are still nowhere near sufficient to make the model acceptable. On the other hand, the ISCO and mass values remain stuck at the lower limit. Finally, by incorporating the magnetic field along with the Lorentz violation, we were able to obtain a statistically consistent model, demonstrating that the combined effect of these parameters is crucial for achieving a satisfactory result. Furthermore, the values for ISCO and mass move away from the lower limit and take on more favorable values.

\begin{table*}[t]
	\centering
	\begin{tabular}{|l|l|l|l|l|l|l|}
		\hline
		Parameter & Best Value & Average $\pm{\sigma}$ & Median & $1{\sigma}$ & $2{\sigma}$ & $3{\sigma}$ \\
		\hline
		$l=\mathcal{B}=0$ & & & & & ${\Delta}AIC=818.02$ & ${\Delta}BIC=820.64$ \\
		\hline
		$M/M_{\odot}$ & $8.5000$ & $8.5060\pm 0.0059$ & $8.5042$ & $\left[8.5010,8.5110\right]$ & $\left[8.5001,8.5225\right]$ & $\left[8.5000,8.5387\right]$ \\
		$r/r_{ISCO}$ & $1.0833$ & $1.0855\pm 0.0019$ & $1.0849$ & $\left[1.0837,1.0872\right]$ & $\left[1.0834,1.0907\right]$ & $\left[1.0833,1.0952\right]$ \\
		Statistics & AIC: $826.02$ & BIC: $823.41$ & LogLik: $-411.01$ & Nº Samples: $500,000$ & Nº Parameters: $2$ & $r_{min}=1.08333$ \\
		\hline
		$\mathcal{B}=0$ & & & & & ${\Delta}AIC=240.19$ & ${\Delta}BIC=241.49$ \\
		\hline
		$l$ & $0.3618$ & $0.3609\pm 0.0143$ & $0.3609$ & $\left[0.3467,0.3752\right]$ & $\left[0.3323,0.3891\right]$ & $\left[0.3179,0.4025\right]$ \\
		$M/M_{\odot}$ & $8.5000$ & $8.5074\pm 0.0073$ & $8.5052$ & $\left[8.5013,8.5136\right]$ & $\left[8.5002,8.5275\right]$ & $\left[8.5000,8.5491\right]$ \\
		$r/r_{ISCO}$ & $1.0833$ & $1.0839\pm 0.0006$ & $1.0838$ & $\left[1.0834,1.0844\right]$ & $\left[1.0833,1.0856\right]$ & $\left[1.0833,1.0873\right]$ \\
		Statistics & AIC: $248.19$ & BIC: $244.27$ & LogLik: $-121.09$ & Nº Samples $500,000$ & Nº Parameters: $3$ & $r_{min}=1.08333$ \\
		\hline
		$l$ and $\mathcal{B}$ free & & & & & ${\Delta}AIC=0.00$ & ${\Delta}BIC=0.00$ \\
		\hline
		$l$ & $0.1002$ & $-0.0909\pm 0.1436$ & $-0.1274$ & $\left[-0.2327,0.0762\right]$ & $\left[-0.2767,0.2240\right]$ & $\left[-0.3145,0.2829\right]$ \\
		$\mathcal{B}$ & $0.0340$ & $0.1185\pm 0.1058$ & $0.0735$ & $\left[0.0364,0.2267\right]$ & $\left[0.0219,0.4159\right]$ & $\left[0.0158,0.4937\right]$ \\
		$M/M_{\odot}$ & $9.1154$ & $9.1823\pm 0.2870$ & $9.1777$ & $\left[8.8860,9.4734\right]$ & $\left[8.6376,9.7695\right]$ & $\left[8.5131,10.0929\right]$ \\
		$r/r_{ISCO}$ & $1.2592$ & $1.4537\pm 0.1521$ & $1.4800$ & $\left[1.2802,1.6086\right]$ & $\left[1.1339,1.6855\right]$ & $\left[1.0866,1.7536\right]$ \\
		Statistics & AIC: $8.00$ & BIC: $2.77$ & LogLik: $0.00$ & Nº Samples: $500,000$ & Nº parameters: $4$ & $r_{min}=1.08333$ \\
		\hline
	\end{tabular}
	\caption{Results of the MCMC simulation for the object XTE 1550-564.}
	\label{XTEdata1}
\end{table*}

Table \ref{XTEdata1} shows the data obtained from the simulations for the object XTE 1550-564. First, for the Schwarzschild case, we see that the AIC and BIC values are much lower compared to those of the object GRO 1655-40. However, they are not nearly sufficient to make the model compatible with our study. In this case as well, the ISCO is constrained to the lower limit, as is the object's mass. When we include the Lorentz violation parameter, we see a significant improvement in the AIC and BIC values, but they are still insufficient to validate the model. Furthermore, the best values for ISCO and mass remain near the lower limit. Now, with the inclusion of the magnetic field and the Lorentz violation, we see that the variations in AIC and BIC become entirely acceptable, so that the model in this case is consistent and statistically reliable. Furthermore, the values for mass and ISCO fall much more closely within the expected range.

\begin{table*}[t]
	\centering
	\begin{tabular}{|l|l|l|l|l|l|l|}
		\hline
		Parameter & Best Value & Average $\pm{\sigma}$ & Median & $1{\sigma}$ & $2{\sigma}$ & $3{\sigma}$ \\
		\hline
		$l=\mathcal{B}=0$ & & & & & ${\Delta}AIC=62.69$ & ${\Delta}BIC=64.00$ \\
		\hline
		$M/M_{\odot}$ & $9.6000$ & $9.6602\pm 0.0587$ & $9.6426$ & $\left[9.6107,9.7107\right]$ & $\left[9.6015,9.8213\right]$ & $\left[9.6001,9.9792\right]$ \\
		$r/r_{ISCO}$ & $1.2413$ & $1.2372\pm 0.0143$ & $1.2370$ & $\left[1.2229,1.2515\right]$ & $\left[1.2091,1.2661\right]$ & $\left[1.1955,1.2811\right]$ \\
		Statistics & AIC: $68.71$ & BIC: $66.10$ & LogLik: $-32.36$ & Nº Samples: $500,000$ & Nº Parameters: $2$ & $r_{min}=1.08333$ \\
		\hline
		$\mathcal{B}=0$ & & & & & ${\Delta}AIC=0.00$ & ${\Delta}BIC=0.00$ \\
		\hline
		$l$ & $0.2706$ & $0.2458\pm 0.0299$ & $0.2468$ & $\left[0.2152,0.2759\right]$ & $\left[0.1840,0.3025\right]$ & $\left[0.1540,0.3273\right]$ \\
		$M/M_{\odot}$ & $11.6112$ & $10.7535\pm 0.5882$ & $10.7936$ & $\left[10.0608,11.3926\right]$ & $\left[9.6748,11.7552\right]$ & $\left[9.6049,12.0367\right]$ \\
		$r/r_{ISCO}$ & $1.0833$ & $1.1443\pm 0.0414$ & $1.1385$ & $\left[1.0990,1.1926\right]$ & $\left[1.0855,1.2290\right]$ & $\left[1.0835,1.2518\right]$ \\
		Statistics & AIC: $6.02$ & BIC: $2.10$ & LogLik: $-0.01$ & Nº Samples $500,000$ & Nº Parameters: $3$ & $r_{min}=1.08333$ \\
		\hline
		$l$ and $\mathcal{B}$ free & & & & & ${\Delta}AIC=1.98$ & ${\Delta}BIC=0.68$ \\
		\hline
		$l$ & $0.1173$ & $-0.0686\pm 0.1342$ & $-0.0962$ & $\left[-0.1955,0.0788\right]$ & $\left[-0.2798,0.2406\right]$ & $\left[-0.3586,0.3124\right]$ \\
		$\mathcal{B}$ & $0.0305$ & $0.1308\pm 0.1117$ & $0.0885$ & $\left[0.0350,0.2494\right]$ & $\left[0.0144,0.4294\right]$ & $\left[0.0021,0.4944\right]$ \\
		$M/M_{\odot}$ & $14.4280$ & $15.8174\pm 1.9413$ & $15.9048$ & $\left[13.8893,17.7310\right]$ & $\left[11.5895,19.5062\right]$ & $\left[9.9523,21.2815\right]$ \\
		$r/r_{ISCO}$ & $1.2689$ & $1.4089\pm 0.1402$ & $1.4078$ & $\left[1.2640,1.5498\right]$ & $\left[1.1354,1.6961\right]$ & $\left[1.0887,1.8474\right]$ \\
		Statistics & AIC: $8.00$ & BIC: $2.78$ & LogLik: $0.00$ & Nº Samples: $500,000$ & Nº parameters: $4$ & $r_{min}=1.08333$ \\
		\hline
	\end{tabular}
	\caption{Results of the MCMC simulation for the object GRS 1915+105.}
	\label{GRSdata1}
\end{table*}

Finally, Table \ref{GRSdata1} presents the data for the object GRS 1915+105. To begin with, in the case of Schwarzschild, we can see that the values obtained for AIC and BIC are much lower than those for the other two objects; even so, these differences do not yet make this model acceptable. On the other hand, although the mass is constrained to the lower limit, the ISCO value obtained in this case is optimal and is not constrained to the lower limit. With the inclusion of the Lorentz violation parameter, we observe significant changes in the results obtained. First, the mass takes on a value further from the lower limit and much more acceptable for our modeling, even though the ISCO value remained at the lower limit. Most importantly, the AIC and BIC values were the lowest and best values, which made their variations ideal for the model's reliability, demonstrating that even a small change in the geometry of spacetime is sufficient to model the frequencies of QPOs. Finally, by incorporating both the Lorentz violation and magnetic field parameters, we were able to obtain the best estimates for the object's mass and ISCO. On the other hand, the AIC and BIC values are slightly higher, but their variations are still within a range that makes the model statistically reliable.

\section{Conclusion}\label{sec5}

In Schwarzschild spacetime, the trajectories of charged particles are quite stable and well-behaved, given that the metric is purely static and spherically symmetric; consequently, the paths taken by the particles are influenced primarily by gravity. For this reason, it is useful to test different variations based on the Schwarzschild metric to see how such changes affect the behaviour of particles around an event horizon. A first variation is the presence of a uniform external magnetic field around the black hole. As we have seen, the presence of this field significantly affects the trajectories of charged particles, exerting an influence comparable to that of the gravitational field. Furthermore, it has been possible to observe varied and distinct behaviours in the trajectories, including behaviours that are not observable in the absence of the magnetic field. On the other hand, we have the Lorentz violation parameter, which is characteristic of the Kalb--Ramond field. In previous work~\cite{Ednaldo2024}, we have seen how the presence of this parameter in the Schwarzschild metric causes certain alterations in spacetime, which consequently affect the trajectories of particles near the event horizon.

It was on the basis of these principles that, in this study, we decided to apply both the Lorentz violation parameter and an external uniform magnetic field, as this allows us to observe how these variables affect the trajectories of charged particles in the vicinity of a black hole. Furthermore, this study provides a framework to model certain astrophysical objects. As we have seen in equations \eqref{phidot}, \eqref{thetaddot2}, and \eqref{rddot2}, the inclusion of the Lorentz violation parameter leads to certain changes in the equations of motion, which were already affected by the presence of the magnetic field term. This results in significant changes to the trajectories and to the conserved quantities, namely energy and angular momentum. Another aspect affected is the innermost stable circular orbit, which, depending on the value of $l$, may settle closer to or slightly further from the event horizon.

Another interesting point worth noting concerns the trajectories obtained and the cases involving energy boundaries. As shown in Figures \ref{1caso} through \ref{2caso3Dl+}, we present representations of the four cases of energy boundaries viewed from the $xz$ and $xy$ planes, along with a three-dimensional visualization. As previously mentioned, the Schwarzschild model with an external magnetic field was used as a reference for these cases. What we can conclude is that the inclusion of the Lorentz violation parameter $l$ completely alters, in some examples, the energy boundary, even changing the type of boundary. This is a very interesting result, as we see that changes in boundary types are not exclusive to the presence of the magnetic field and its change in magnitude.

The fundamental frequencies ${\omega}_{\phi}$, ${\omega}_{\theta}$, and ${\omega}_{r}$ yielded interesting results in our tests involving the magnetic field and Lorentz violation. The frequency ${\omega}_{\phi}$ underwent significant changes in amplitude when the parameters $\mathcal{B}$ and $l$ were altered. The frequency ${\omega}_{\theta}$ exhibited a modest dependence on $\mathcal{B}$ and $l$, varying only weakly for the parameter range considered, consistent with the analytical expression ${\partial}^{2}_{\theta}V_{\mathrm{eff}}|_{{\theta}={\pi}/2}=2f(r)(\mathcal{L}^{2}/r^{2}-\mathcal{B}^{2}r^{2})$. On the other hand, ${\omega}_{r}$ exhibited abrupt changes in intensity, even with the slightest variations in the parameters.

The results of the MCMC analysis were also promising. Based on the values of ${\Delta}AIC$ and ${\Delta}BIC$, and LogLik (which measures the likelihood), we can determine which models are statistically reliable for application to the astrophysical objects in our study when modeling quasi-periodic oscillations. As is well known, a pure Schwarzschild model is not capable of modeling these objects, and the mere presence of a magnetic field already provides a minimally satisfactory model. In Tables \ref{GROdata1} and \ref{XTEdata1} for the objects GRO 1655-40 and XTE 1550-564, respectively, we observe that it is only possible to fully model the QPOs by combining the magnetic field with Lorentz violation. On the other hand, in Table \ref{GRSdata1}, which contains data for the object GRS 1915+105, we see that the Lorentz violation parameter alone provides an excellent model for the quasi-periodic oscillations. This is a surprising result, as it shows that a modification of the spacetime geometry alone can provide the appropriate model for objects with sufficiently large mass.

More specifically, for the Schwarzschild case, all three objects had their best-fit mass values at the observational lower limits and $r/r_{ISCO}$ at the prior limit of $1.0833$, with $\Delta AIC$ and $\Delta BIC$ values far above the threshold for statistical reliability ($\Delta AIC \ge 10$ indicating strong evidence against the model). When the Lorentz violation parameter $l$ was introduced while keeping $\mathcal{B}=0$, the best-fit values for mass and ISCO remained at the limits for GRO 1655-40 and XTE 1550-564, despite improvements in the information criteria. Only GRS 1915+105 showed a substantial improvement, with the best-fit mass rising to $M/M_{\odot}=11.6112$ and the information criteria dropping to $\Delta AIC = \Delta BIC = 0$, making the model statistically reliable.

Finally, with both $\mathcal{B}$ and $l$ free, all three objects achieved statistically consistent models. For GRO 1655-40, the best-fit parameters were $\mathcal{B}=0.0449$ and $l=0.0833$, with $M/M_{\odot}=6.2970$ and $r/r_{ISCO}=1.2073$. For XTE 1550-564, $\mathcal{B}=0.0340$ and $l=0.1002$, with $M/M_{\odot}=9.1154$ and $r/r_{ISCO}=1.2592$. For GRS 1915+105, $\mathcal{B}=0.0305$ and $l=0.1173$, with $M/M_{\odot}=14.4280$ and $r/r_{ISCO}=1.2689$, yielding $\Delta AIC=1.98$ and $\Delta BIC=0.68$, which are well within the range of statistical equivalence.

Based on the results obtained from the MCMC statistics, we can conclude that the association between the magnetic field and the parameter $l$ means that the value of this parameter does not need to be high, and, consequently, the magnetic field strength does not need to be strong either. This can be seen by comparing the cases with $\mathcal{B}=0$ and $l$ free against those with both $\mathcal{B}$ and $l$ free in Tables \ref{GROdata1}, \ref{XTEdata1}, and \ref{GRSdata1}. Furthermore, it is notable that as the object's mass increases, the values of $l$ decrease. From the data presented in the tables, it is evident that the greater the mass of the object, the smaller the Lorentz violation parameter $l$ must be to model the quasi-periodic oscillations. This suggests a mass-dependent scaling of the KR parameter.

The results presented in this work open several avenues for future investigation. First, the mass-dependent scaling of the Lorentz violation parameter observed in the MCMC analysis warrants a deeper theoretical understanding: a systematic study across a broader sample of astrophysical black holes, including both stellar-mass and supermassive candidates, could reveal whether this trend is a generic feature of the KR model or specific to the objects considered here. Second, the analysis could be extended to rotating black hole solutions in the presence of the Kalb--Ramond field, where the interplay between spin, the Lorentz violation parameter, and the external magnetic field would introduce richer dynamics and additional observable signatures, such as modifications to the black hole shadow and jet power. Third, the inclusion of more sophisticated disk models, beyond the test-particle approximation employed in this work, would allow for a more realistic description of the accretion flow and its radiative properties, potentially linking the KR parameter to spectral features observed by X-ray missions such as \textit{NICER} and \textit{XRISM}. Fourth, gravitational wave observations from extreme-mass-ratio inspirals (EMRIs), to be probed by the future LISA mission, could provide independent and highly sensitive constraints on the Lorentz violation parameter through the dephasing of the gravitational waveform induced by deviations from the Kerr geometry. Finally, the formalism developed here could be applied to other Lorentz-violating gravity models, such as Einstein--\AE{}ther theory and Bumblebee gravity, allowing for a comparative study of their predictions against QPO data and other astrophysical observables. Collectively, these directions would deepen our understanding of the observable consequences of Lorentz violation in strong-field astrophysical environments and bring us closer to constraining or detecting quantum-gravity signatures with current and next-generation instruments.

In summary, this work has demonstrated that the combined effect of a Kalb--Ramond Lorentz violation parameter and an external uniform magnetic field provides a viable and statistically robust framework for modeling the quasi-periodic oscillations observed in microquasars. The ability of the KR parameter alone to model QPOs in the most massive object of our sample, GRS 1915+105, is particularly noteworthy and suggests that Lorentz-violating modifications to the spacetime geometry may leave observable imprints in the strong-field regime. The small values of both $\mathcal{B}$ and $l$ required to achieve excellent fits highlight the sensitivity of QPO frequencies to subtle changes in the underlying gravitational theory, reinforcing their role as powerful probes of fundamental physics.

\section*{Acknowledgments}

MER thanks Conselho Nacional de Desenvolvimento Científico e Tecnológico - CNPq, Brazil, for partial financial support. This study was financed in part by the Coordenação de Aperfeiçoamento de Pessoal de Nível Superior - CAPES, Brazil - Finance Code 001. FSNL acknowledges support from the Fundação para a Ciência e a Tecnologia (FCT) Scientific Employment Stimulus contract with reference CEECINST/00032/2018, and funding through the research grant UID/04434/2025.



\end{document}